%% file: compiler-bugs.tex
\documentclass[acmsmall]{acmart}
\setcopyright{none}

\usepackage{booktabs}   
\usepackage{subcaption} 

\usepackage{listings}
\usepackage{multirow}
\usepackage{xspace}
\usepackage{todonotes}
\usepackage{enumitem}

\newcommand{\myparagraph}[1]{\smallskip \noindent{\bf #1}}

\begin{document}

\title{A Systematic Impact Study for Fuzzer-Found Compiler Bugs}         


\author{Micha\"el Marcozzi}
\authornote{Micha\"el Marcozzi and Qiyi Tang have contributed equally to the presented experimental study.}          
\author{Qiyi Tang}\authornotemark[1]
 \author{Alastair F. Donaldson}  
\author{Cristian Cadar}
\affiliation{
  \institution{~\\Imperial College London}            
  \city{London}
  \country{United Kingdom}                    
}
\email{michael.marcozzi@gmail.com}          
\email{qiyi.tang71@gmail.com}         
\email{c.cadar@imperial.ac.uk}          
\email{alastair.donaldson@imperial.ac.uk}   


\input{macros}

\input{data}

\input{0abstract.tex}




\maketitle

\section{Introduction}
\label{sec:intro}
\input{1introduction}

\section{Background}
\label{sec:background}
\input{2background}

\section{Methodology}
\label{sec:methodology}
\input{3approach}

\section{Experimental setup}
\label{sec:setup}
\input{4infrastructure}

\section{Results}
\label{sec:results}
\input{5results}

\section{Threats to validity}
\label{sec:threats}
\input{6threats}

\section{Future Work}
\label{sec:future}
\input{future}

\section{Related Work}
\label{sec:related}
\input{7related}

\section{Conclusion}
\label{sec:conclusion}
\input{8conclusion}

\section{Appendices}
\label{sec:appendix}
\input{appendix}

\input{9acknowledgment}

\newpage
\bibliography{srg-bib/cadar-macros,srg-bib/cadar,srg-bib/cadar-crossrefs,compiler-bugs}

\end{document}

%% file: macros.tex
\newcommand{\eg}{e.g.~}
\newcommand{\ie}{i.e.~}
\newcommand{\etc}{etc.~}
\newcommand{\etal}{et al.~}

\newcommand{\OTHERSPACEHACK}[1]{\vspace{#1}}
\newcommand{\SPACEHACK}[1]{} 
\newcommand{\ADComment}[1]{\textcolor{red}{[AD: #1]}}
\newcommand{\CC}[1]{\textcolor{blue}{[CC: #1]}}
\newcommand{\MM}[1]{\textcolor{orange}{[MM: #1]}}
\definecolor{OliveGreen}{rgb}{0,0.6,0}
\newcommand{\QT}[1]{\textcolor{OliveGreen}{[QT: #1]}}


\newcommand{\apache}{\textit{Apache}\xspace}
\newcommand{\grep}{\textit{Grep}\xspace}
\newcommand{\openssh}{\textit{OpenSSH}\xspace}
\newcommand{\samba}{\textit{Samba}\xspace}
\newcommand{\simplebuild}{\textit{Simple Build}\xspace}
\newcommand{\sqlite}{\textit{SQLite}\xspace}
\newcommand{\zsh}{\textit{zsh}\xspace}
\newcommand{\leveldb}{\textit{leveldb}\xspace}
\newcommand{\modgearman}{\textit{mod-gearman}\xspace}
\newcommand{\phyml}{\textit{phyml}\xspace}
\newcommand{\libgthreed}{\textit{libg3d}\xspace}

%% file: data.tex
\newcommand{\numPackages}[0]{309\xspace}

\newcommand{\numFuzzerBugs}{27\xspace}
\newcommand{\numEMIBugs}{10\xspace}
\newcommand{\numCsmithBugs}{10\xspace}
\newcommand{\numFuzzerAndAliveBugs}{35\xspace}
\newcommand{\numUserBugs}{10\xspace}
\newcommand{\numAliveBugs}{8\xspace}
\newcommand{\numNonFuzzerBugs}{18\xspace}
\newcommand{\numAllBugs}{45\xspace}
\newcommand{\numFailedBuildPackages}{17\xspace}

\newcommand{\numPatchReachedFuzzers}{27\xspace}
\newcommand{\numPatchReachedAlive}{7\xspace}
\newcommand{\numPatchReachedUser}{8\xspace}

\newcommand{\numPatchNotReachedAlive}{1\xspace}
\newcommand{\numPatchNotReachedUser}{2\xspace}

\newcommand{\numMaybeTriggeredFuzzerBugs}{22\xspace}
\newcommand{\numNeverTriggeredCsmithEMIBugs}{2\xspace} 
\newcommand{\numMaybeTriggeredAliveBugs}{2\xspace}
\newcommand{\numNeverTriggeredAliveBugs}{6\xspace} 

\newcommand{\numMaybeTriggeredUserBugs}{4\xspace}

\newcommand{\numDefinitelyTriggeredFuzzerBugs}{1\xspace}
\newcommand{\numDefinitelyTriggeredAliveBugs}{0\xspace}
\newcommand{\numDefinitelyTriggeredUserBugs}{1\xspace}

\newcommand{\numBinaryDiffsFuzzerBugs}{12\xspace}
\newcommand{\numBinaryDiffsEMI}{4\xspace}
\newcommand{\numBinaryDiffsCsmith}{7\xspace}
\newcommand{\numBinaryDiffsUserBugs}{2\xspace}
\newcommand{\numBinaryDiffsAliveBugs}{2\xspace}

\newcommand{\percentPropagatedFuzzerBugs}{21\%\xspace} 
\newcommand{\percentPropagatedUserBugs}{9\%\xspace} 
\newcommand{\percentPropagatedAliveBugs}{53\%\xspace} 

\newcommand{\percentVelvetFailure}{52\%\xspace} 
\newcommand{\numPatchReachMoreThanHalfFuzzerBugs}{19\xspace}
\newcommand{\numNeverReachFuzzerBugs}{0\xspace}
\newcommand{\percentTriggerByReachFuzzerBugs}{39\%\xspace} 
\newcommand{\percentTriggerByBuildFuzzerBugs}{28\%\xspace} 
\newcommand{\percentTriggerByBuildPreciseFuzzerBugs}{13\%\xspace} 
\newcommand{\percentBinDiffByBuildFuzzerBugs}{6\%\xspace} 
\newcommand{\percentTriggerEMI}{31\%\xspace} %
\newcommand{\percentTriggerCsmith}{34\%\xspace} %
\newcommand{\percentBinDiffEMI}{5\%\xspace} %
\newcommand{\percentBinDiffCsmith}{10\%\xspace} %
\newcommand{\percentReachEMI}{79\%\xspace} %
\newcommand{\percentReachCsmith}{81\%\xspace} %

\newcommand{\percentTSFailFuzzers}{13\%\xspace} 
\newcommand{\percentBinDiffByBuildUser}{2\%\xspace} 
\newcommand{\numBinDiffCsmithEMI}{11\xspace}  
\newcommand{\percentBinDiffByBuildCsmithEMI}{8\%\xspace} 
\newcommand{\numLOC}{10M\xspace} 
\newcommand{\numLOCMillion}{10\xspace}

%% file: 0abstract.tex

\begin{abstract}

  Despite much recent interest in randomised testing
  (fuzzing) of compilers, the practical impact of fuzzer-found compiler bugs on
  real-world applications has barely been assessed.  We present the
  first quantitative and qualitative study of the tangible impact of
  miscompilation bugs in a mature compiler. We follow a rigorous
  methodology where the bug impact over the compiled application is
  evaluated based on (1) whether the bug appears to trigger during
  compilation; (2) the extent to which generated assembly code changes
  syntactically due to triggering of the bug; and (3) whether such
  changes cause regression test suite failures, or whether we can manually find
  application inputs that trigger execution divergence due to such changes.
  The study is
  conducted with respect to the compilation of more than
  \numLOCMillion million lines of C/C++ code from \numPackages Debian
  packages, using 12\% of the historical and now fixed miscompilation
  bugs found by four state-of-the-art fuzzers in the Clang/LLVM
  compiler, as well as 18 bugs found by human users compiling real
  code or as a by-product of formal verification efforts.
  The results show that almost half of the fuzzer-found bugs propagate to the generated binaries for at least one package, in which case only a very small part of the binary is typically affected, yet causing two failures when running the test suites of all the impacted packages. User-reported and formal
  verification bugs do not exhibit a higher impact, with a lower rate of triggered bugs and one test failure. The manual analysis of a selection of the syntactic changes
 caused by some of our bugs (fuzzer-found and non fuzzer-found) in package assembly code, shows that either these changes have no semantic impact or that they would require very specific runtime circumstances to trigger execution divergence.

  %

\end{abstract}

%% file: 1introduction.tex

\myparagraph{Context.} Compilers are among the most central components
in the software development toolchain. While software developers often
rely on compilers with blind confidence, bugs in state-of-the-art
compilers are frequent \cite{compiler-bugs:issta16}; for example,
hundreds of bugs in the Clang/LLVM and GCC compilers are fixed each
month.
The consequence of a functional
compiler bug may be a compile-time crash or a \textit{miscompilation},
where wrong target code is silently generated. While compiler crashes
are spotted as soon as they occur, miscompilations can go unnoticed
until the compiled application fails in production, with potentially
serious consequences.  Automated compiler test generation has been a
topic of interest for many years
(see \eg~\cite{cobol-testing62,Hanford70,Purdom72,Boujarwah97,Wichmann98,Kossatchev2005}),
and recent years have seen the development of several \textit{compiler
fuzzing} tools that employ randomised testing to search for bugs in
(primarily C) compilers~\cite{csmith,emi,orange4,yarpgen}. Such \emph{compiler fuzzers} have practically demonstrated their ability to
find hundreds of bugs, and particularly many miscompilations, in widely-used
compilers.

\myparagraph{Problem.} In our experience working in the area \cite{OpenGLtesting:met16,compiler-bugs:PACMPL17,clsmith}, we have found
compiler fuzzing to be a contentious topic.  Research talks on compiler fuzzing
are often followed by questions about the importance of the discovered bugs, and whether compiler fuzzers might be improved by taking inspiration from bugs encountered by users of compilers ``in the wild''. Some (including some reviewers of this paper) argue that any miscompilation bug,
whether fuzzer-found or not, is a ticking bomb that should be regarded as severe, or avoided completely via formal verification (in the spirit of CompCert~\cite{CompCert}).  Others (see e.g.~\cite{anti-comp-fuzzing15}) question whether the bugs
found by compiler fuzzers are worth fixing.
In the context of fuzzing a mature compiler used for processing \emph{non-critical} software, where a level of unreliability can be tolerated in exchange for reduced development costs, it is reasonable to question the importance of
miscompilation bugs found by compiler fuzzers.
First, there is an argument that a combination of regular testing of the compiler by its developers and its intensive operation by end-users is likely to flag up the serious bugs as they arise. It is thus unclear whether this leaves enough space in practice for
fuzzers to find bugs of more than mild importance.  Second, by their very nature, fuzzers detect miscompilations via
\textit{artificial} programs, obtained by random
generation of code from scratch or random modification of existing
code.  It is thus unclear whether the code patterns that trigger
these miscompilations are likely to be exhibited by applications in the
wild.
The lack of a fine-grained and quantitative study of the practical
impact of miscompilation bugs in a mature compiler means that we have
little but anecdotal evidence to support or rebut these
points.\footnote{A recent empirical study of GCC and LLVM compiler
bugs~\cite{compiler-bugs:issta16} provides numerous insights into bug
characteristics, but does not address the extent to which bugs found
by fuzzers impact on code found in the wild.} More practically, since
the bug databases of popular mature compilers such as Clang/LLVM have
thousands of open bug reports, such a study could also provide
insights into the risk associated with leaving bugs unfixed for long periods
of time,
as well as establishing whether the way compiler developers
prioritise bugs for fixing appears to correlate with their impact.

\myparagraph{Objectives and Challenges.} Our main objectives are (a)~to
study how much the miscompilation bugs found in a widely-used compiler
affect the compilation of real-world applications, and (b)~to compare
the impact of bugs found by fuzzers (using
artificial programs) and by other means (notably via compilation of real
code). We have approached this study with an open mind, not wishing to
specifically show that fuzzer-found bugs do or do not matter, but
rather (1)~to take sensible steps towards a methodology to assess and
compare the impact of miscompilation bugs and (2)~to apply this
methodology to a significant but tractable sample of bugs in a mature
compiler, over a feasible range of real-world applications.

\myparagraph{Methodology.} As a simplifying assumption, we choose to
liken the impact of a miscompilation bug on an application to its
ability to change the semantics of the application.\footnote{We do not consider here the impact over the non-functional properties of the application, such as security or performance.}
We take advantage of the large number of miscompilation bugs in
open-source compilers that have been publicly reported and fixed.
Given a fixed bug, we analyse the fix and devise a change to the
compiler's code that emits a warning when the faulty code is reached
and a local faulty behaviour is triggered.  The warnings issued when
compiling the application with the modified compiler provide
information on whether the bug is reached and triggered at all.  In
cases where warnings are issued, we compare the application binaries
produced by the faulty and the fixed compiler versions (subject to
appropriate care to ensure reproducible builds), counting the number
of functions that exhibit bug-induced syntactic differences.  If such
differences are detected, we run the application's standard test suite
and look for discrepancies caused by these syntactic differences at
runtime.  In cases where no such discrepancies are detected, we may
also inspect manually the application's buggy and fixed binaries and
try to craft specific inputs to trigger a runtime divergence when
executing the syntactically different sections of these binaries. The
frequency of the syntactic differences between the two binaries and
their ability to trigger test discrepancies or runtime divergences
shed light on the impact of the bug on the application's correctness
and reliability.

\myparagraph{Experimental Study.} We present a rigorous
study of the real-world impact of miscompilation bugs in a mature
compiler over a large set of diverse applications. In particular, we
sample a set of 12\% of the fixed miscompilation bugs detected by
four state-of-the-art families of fuzzers targeting the Clang/LLVM C/C++
compiler: Csmith~\cite{csmith,creduce,chen2013},
EMI~\cite{emi,athena,le2015issta}, Orange3/4~\cite{orange3,orange4}
and Yarpgen~\cite{yarpgen}. We also select a set of \numUserBugs bugs
detected directly by end users of Clang/LLVM when compiling their
applications, and consider the \numAliveBugs bugs found as a
by-product of applying the Alive formal verification
tool~\cite{alive}. 
Then, following our methodology, we evaluate the impact of all these
bugs when compiling \numPackages Debian packages, such
as \apache, \grep and \samba, not used by the Clang/LLVM developers to test
regularly the compiler. Finally, we compare the impact of the
fuzzer-found bugs with the impact of the user-reported and Alive bugs.

\myparagraph{Contributions.} To sum up, our main contributions are:

\begin{enumerate}

\item A three-stage methodology (detailed in
  \S\ref{sec:methodology}) for assessing the impact of a
  miscompilation bug on the correctness and reliability of a given application (a)~during compilation (is
  the faulty compiler code reached and triggered?), (b)~by syntactically
  analysing the generated code (how much does the fault impact the
  assembly syntax of the resulting binary?)  and (c) by dynamically
  probing the produced binary (can the fault lead to observable differences in the runtime behaviour of the application?)

\item The first systematic study on the real-world
  impact of miscompilation bugs in a mature compiler (presented
  in \S\ref{sec:setup}, \S\ref{sec:results} and \S\ref{sec:threats}),
  which applies our methodology (a) to evaluate the impact of a sample
  of \numAllBugs fixed miscompilation bugs in Clang/LLVM
  over \numPackages diverse Debian packages totalling \numLOC lines of
  C/C++, and (b) to compare the impact of \numFuzzerBugs of these bugs
  that are fuzzer-found with the impact of \numNonFuzzerBugs of these
  bugs found either by users compiling real code or by the Alive tool,
  over the same Debian packages. The study required a huge amount of
  manual effort and five months of machine time.

\end{enumerate}

\myparagraph{Summary of Main Findings.}  Our top-level findings include
that the code associated with each fuzzer-found bug is reached always at least once and typically quite frequently
when compiling our set of real-world applications, that our bug
conditions trigger, and almost half of the bugs result in binary-level
differences for some packages.  However, these differences affect a
tiny fraction of the functions defined in the application's code and
they cause a couple of test failures in total. Regarding the user-reported and Alive bugs, there exist bugs for which the associated code is {never} reached during the compilation of our set of packages. In addition, user-reported and Alive bugs have their conditions
triggered less frequently on average, and lead to a single test failure. The manual analysis of a selection of the binary-level
differences caused by some of our compiler bugs,
both fuzzer-found and non fuzzer-found, shows that either these binary differences have no impact over the semantics of the application, or that they would require very specific runtime circumstances in order to trigger execution divergence.

Considering the two aforementioned objectives of this work, our
results support the argument that (a) miscompilation bugs in a
generally high-quality, widely-used production compiler seldom impact
the correctness and reliability of deployed applications,
as encoded by their test suites and (in a few cases) based on our manual
analysis, and that (b) bugs in a mature compiler discovered using a
fuzzer relying on artificial code appear to have at least as much
impact as those found via other sources, notably bugs reported by the
compiler users because they were affecting real code.

\myparagraph{Future Work.} We discuss (\S\ref{sec:future}) further
directions to evaluate even better the impact of miscompilation bugs,
for example by considering their effect over the application's non-functional
properties (\eg security or performance) or by using symbolic
execution~\cite{symex:cacm} to trigger miscompilation-caused runtime
failures.

\myparagraph{Artifact.} Our experimental data and infrastructure are
available as a part of an associated
artifact.\footnote{\url{https://srg.doc.ic.ac.uk/projects/compiler-bugs/}}

%% file: 2background.tex

We make precise relevant details of {compilation} faults and
{miscompilation} failures (\S\ref{sec:background:miscompilations}), provide an overview of
the compiler validation tools considered in the study (\S\ref{sec:background:fuzzers}),
provide relevant background on Clang/LLVM, including
details of the bug tracker from which we have extracted compiler bug
reports (\S\ref{sec:background:clangllvm}), and describe the
standard framework used to build and test Debian packages
(\S\ref{sec:background:debian}), which is at the core of the
experimental infrastructure deployed for the study.

\subsection{Compilation Faults and Miscompilation Failures}\label{sec:background:miscompilations}

We describe the notion of a miscompilation bug using a simple but
pragmatic model of how a compiler works.
A modern compiler typically converts a source program into an
intermediate representation (IR), runs a series of {passes} that
transform this IR, and finally emits target code for a specific
architecture.
For source program $P$ and input $x$, let $P(x)$ denote the set of
results that $P$ may produce, according to the semantics of the
programming language in question.  This set may have multiple values,
for example if $P$ can exhibit nondeterminism.  Similarly, let $T(x)$ denote the
set of results that target program $T$ may produce, according to the
semantics of the target architecture.  For ease of presentation we
assume that source and target programs always terminate, so that
$P(x)$ and $T(x)$ cannot be empty.  A compilation
$P\rightarrow T$ 
is \emph{correct} with respect to an input $x$ if $T$ is a \emph{refinement} of $P$, i.e. if $T(x)
\subseteq P(x)$.  That is, the target program respects the semantics
of the source program: any result that the target program may produce
is a result that the source program may also produce.  Otherwise the
compilation exhibits a \emph{miscompilation failure} with respect to
input $x$. We call a \textit{compilation fault} any internal compiler operation which is incorrect during a compilation.
A miscompilation failure is always caused by a compilation fault, but
a compilation can exhibit a fault and no miscompilation failure if,
for example, the fault makes the compiler crash or only impacts IR
code that will be detected as dead and removed by a later pass.  In
such a case, the fault is said not to \emph{propagate} to a
miscompilation failure.

\subsection{Fuzzers and Compiler Verification Tools Studied}\label{sec:background:fuzzers}

Our study focuses on four compiler fuzzing tools (in two cases the
``tool'' is actually a collection of closely-related tools), chosen
because they target C/C++ compilers and have found bugs
in recent versions of the Clang/LLVM compiler framework.  We also
consider some bugs found via application of the Alive tool for
verification of LLVM peephole optimisations.  We briefly summarise
each tool, discussing related work on compiler validation more broadly
in \S\ref{sec:related}.

\myparagraph{Csmith.} The Csmith tool~\cite{csmith} randomly generates C
programs that are guaranteed to be free from undefined and unspecified
behaviour.  These programs can then be used for \emph{differential
  testing}~\cite{mckeeman:diff-test} of multiple compilers that agree
on implementation-defined behaviour: discrepancies between compilers
indicate that miscompilations have occurred.
By November 2013 Csmith had been used to find and report 481 bugs in
compilers including LLVM, GCC, CompCert and suncc, out of which about
120 were miscompilations.\footnote{\url{https://github.com/csmith-project/csmith/blob/master/BUGS_REPORTED.TXT}}
Many bugs subsequently discovered using Csmith are available from the bug trackers of the targeted
compilers.

\myparagraph{EMI.} The Orion tool~\cite{emi} introduced the idea of
\emph{equivalence modulo inputs} (EMI) testing.  Given a deterministic C program $P$ (\eg an existing application or a program generated by Csmith) together
with an input $x$ that does not lead to undefined/unspecified
behaviour, Orion profiles the program to find those statements that
are \emph{not} covered when the program is executed on input $x$.  A
set of programs are then generated from $P$ by randomly deleting such
statements.  While very different from $P$ in general, each such
program should behave functionally identically to $P$ when executed on
input $x$; discrepancies indicate miscompilations.  Follow-on tools,
Athena~\cite{athena} and Hermes~\cite{hermes}, extend the EMI idea using
more advanced profiling and mutation techniques; we refer to the three tools collectively as EMI. To date, the project has enabled the discovery of more than 1,600
bugs in LLVM and GCC, of which about 550 are miscompilations.\footnote{\url{http://web.cs.ucdavis.edu/~su/emi-project}}

\myparagraph{Orange3/4.} The Orange3~\cite{orange3} and Orange4~\cite{orange4} tools
can be used to fuzz C compilers via a subset of the C language,
focussing primarily on testing compilation of arithmetic expressions.
Orange3 generates a program randomly, keeping track during generation
of the precise result that the program should compute.  Orange4
instead is based on transforming a test program into equivalent forms
that are all guaranteed to generate the same output if compiled correctly.
Transformations include
adding statements to the program or expanding constants into
expressions.
The tools have led to the reporting of 60 bugs in LLVM and GCC, out of which
about 25 are miscompilations.\footnote{\url{https://ist.ksc.kwansei.ac.jp/~ishiura/pub/randomtest/index.html}}

\myparagraph{Yarpgen.}
The Intel-developed ``Yet Another Random Program Generator'' (Yarpgen)
tool~\cite{yarpgen} takes a Csmith-like approach to generating random
programs.  It accurately detects and avoids undefined behaviour by
tracking variable types, alignments and value ranges during program
generation.  It also incorporates policies that guide random
generation so that optimisations are more likely to be applied to the
generated programs.
It has been used to report more than 150 bugs in LLVM and GCC, out of
which 38 are miscompilations.\footnote{\url{https://github.com/intel/yarpgen/blob/master/bugs.rst}} 

\myparagraph{Alive.} The Alive project~\cite{alive} provides a language to encode formal specifications of LLVM peephole optimisations, together with an SMT-based verifier to either prove them correct or provide counterexamples to correctness.  Once an optimisation specification has been proven correct, Alive can generate LLVM-compatible C++ code that implements the optimisation. More than 300 optimisations were verified in this way, leading to the discovery of 8 miscompilation bugs in LLVM as a by-product.

\subsection{Clang/LLVM Framework}\label{sec:background:clangllvm}

The Clang/LLVM framework~\cite{llvm} is one of the most popular
compiler frameworks, used by a large number of research and commercial
projects. Written in C++, it supports several source languages (e.g.\
C, C++, Objective-C and Fortran) and many target architectures (e.g.\ x86,
x86-64, ARM, PowerPC).
%

The bugs discussed in this paper are analysed in the context of
compiling C/C++ code to x86 machine code.
They were all reported on the Clang/LLVM online bug
tracker.\footnote{\url{https://bugs.llvm.org/}}
A typical bug report includes a fragment of code that demonstrates the
miscompilation, together with the affected Clang/LLVM versions and
configurations (target architectures, optimisation levels, etc.). The
bug is given a unique ID and classified by severity, being either
ranked as an enhancement request, a normal bug or a release blocker. A
public discussion usually follows, involving the compiler developers
who may end up writing a fix for the bug, if judged necessary. The fix
is applied (with attached explanatory comments) directly within the
public Clang/LLVM SVN
repository.\footnote{\url{http://llvm.org/viewvc}}
The revision number(s)
where the fix was applied is typically provided to close the bug
report.

\subsection{Build and Test Framework for Debian Packages }\label{sec:background:debian}

Debian is a well-known open-source operating system and software
environment. It provides a popular repository for compatible packaged
applications,\footnote{\url{https://www.debian.org/distrib/packages}}
together with
a standard framework to facilitate compiling these
packages from source and testing them. The components of the Debian framework that we use in this study
are Simple Build,\footnote{\url{https://wiki.debian.org/sbuild}}
Reproducible
Builds\footnote{\url{https://wiki.debian.org/ReproducibleBuilds}} and
Autopkgtest.\footnote{\url{https://manpages.debian.org/testing/autopkgtest}} 

Simple Build provides a standard way to build any package
from source in a customised and isolated build environment. The
infrastructure provides simple primitives (1) to set up this environment
as a tailored Debian installation within a chroot
jail,\footnote{\url{http://man7.org/linux/man-pages/man2/chroot.2.html}}
(2) to gather packaged sources for any package and compile them within the
environment, and (3) to revert the environment to its initial
state after building a package.

Reproducible Builds is an initiative to drive package developers towards ensuring that identical
binaries are always generated from a given source. This makes it possible to check that no vulnerabilities or backdoors have
been introduced during package compilation, by cross-checking the
binaries produced for a single source by multiple third parties.  The
initiative notably provides a list of the packages for which the build
process is (or can easily be made) reproducible.

Autopkgtest is the standard interface for Debian developers to embed
tests suites in their packages. As soon as the developers provide a
test suite in the requested format, Autopkgtest enables running these
tests over the package in a Debian environment using a single
command. This environment can be the local machine, a distant machine,
or a local installation within a chroot jail or emulator.

%% file: 3approach.tex

Our methodology focuses solely on miscompilation bugs reported in open-source compilers, and for which an associated patch that fixes the bug has been made available.  We refer to this patch as the \emph{fixing patch} for the bug, the version of the compiler just before the patch was applied as the \emph{buggy compiler}, and the version just after the patch was applied as the \emph{fixed compiler}.  

Given a miscompilation bug and fixing patch, together with a set of applications to compile, we evaluate the practical impact of the bug on each application in three successive stages:


\begin{enumerate}

\item \textbf{Compile-time analysis}. We check whether the compiler code affected by the fixing patch is reached when the application is compiled with the fixed compiler.  If so, we check whether the conditions necessary for the bug to trigger in the buggy compiler hold, indicating that a compilation fault occurred.

\item \textbf{Syntactic binary analysis}.
We statically check how much the application binaries generated by the
buggy and fixed compiler are different. More precisely, we count how
many of the functions defined in the produced assembly code have
bug-induced syntactic differences, hypothesising that the larger
the number of these differently-compiled functions the higher the chance that the
compilation fault might impact the application's execution.

\item \textbf{Dynamic binary analysis}. We probe dynamically whether the syntactic differences spotted at Stage~2 can make the binaries generated by the buggy and fixed compiler diverge semantically at runtime. To do so, we run the application's test suite twice, once against each binary. If no differences in the test results are detected, we also manually inspect a sample of the spotted syntactic differences for some applications, with the aim of crafting, if they exist, inputs that would trigger a runtime divergence. Based on this manual inspection, we infer a more general qualitative estimation of the likelihood with which syntactic differences induced by the bug might impact the application's semantics.
\end{enumerate}



We now discuss each of the three stages of our approach in detail, in
the process describing the steps we followed to curate a set of bugs
with associated compiler versions and fixing patches.

\subsection {Stage 1: Compile-time Analysis}
\label{sec:stage-one}

For each bug, the first stage of our approach relies on isolating a fixing patch and preparing appropriate compile-time checks for the conditions under which the
compilation fault would occur.  We accomplished this by careful manual review of
the bug tracker report associated with each miscompilation bug. We limited our attention to bugs where it was clear from discussion
between developers in the bug tracker that the fixing patch was incorporated in a single
revision (or several contiguous revisions) of the compiler sources and in
isolation from any other code modifications.

As a simple running example, the fixing patch for Clang/LLVM bug \#26323 (found by one of the EMI fuzzers)  is Clang/LLVM revision $258904$,\footnote{\url{http://llvm.org/viewvc/llvm-project?view=revision&revision=258904}} which makes the following change:
\begin{lstlisting}[basicstyle=\ttfamily\footnotesize]
  - if (Not.isPowerOf2()) {
  + if (Not.isPowerOf2()
  +     && C->getValue().isPowerOf2()
  +     && Not != C->getValue()) {
         /* CODE TRANSFORMATION */  }
\end{lstlisting}
It is clear from the fixing patch and its explanatory comments on SVN that the bug is fairly simple and localised, fitting a common bug pattern identified by the Csmith authors~\cite[\S3.7]{csmith} where the precondition associated with a code transformation is incorrect.  As a consequence, the transformation can be mistakenly applied, possibly resulting in a miscompilation.  The fix simply strengthens the precondition.

We found and discarded a small number of bugs
whose fixes were applied together with other code modifications and/or via a
series of non-contiguous compiler revisions, except in one case (Clang/LLVM bug \#27903, reported by an end-user) where we found
it straightforward to determine an independent fixing patch for the bug from the
two non-contiguous patches that were used in practice. The first patch, meant as temporary, deactivated the faulty feature triggering the miscompilation, while the second patch permanently fixed this feature and reactivated it. Our independent patch leaves the feature activated and applies the permanent fix of the second patch. 

Having identified a fixing patch and understood the conditions under which the compilation fault would possibly trigger, we
modify the fixed compiler to print warnings (1)~when at least one of the basic blocks affected by the fixing patch is reached during compilation, and (2)~when upon reaching the fixing patch, the conditions under which a compilation fault would possibly have occurred had the patch not been applied are triggered.  In our running example this involves detecting when \lstinline[basicstyle=\ttfamily\small]{Not.isPowerOf2()} holds but \lstinline[basicstyle=\ttfamily\small,breaklines=true]{C->getValue().isPowerOf2()} \lstinline[basicstyle=\ttfamily\small,breaklines=true]{&& Not != C->getValue()} does not.  The fixing patch augmented with warning generation is as follows:
\begin{lstlisting}[basicstyle=\ttfamily\footnotesize]
warn("Fixing patch reached");
if (Not.isPowerOf2()) {
  if (!(C->getValue().isPowerOf2()
      && Not != C->getValue())) {
    warn("Fault possibly triggered");
  } else { /* CODE TRANSFORMATION */  } }
\end{lstlisting}

We sanity-check the correctness of the so crafted \textit{warning-laden compiler} by making sure that the warnings are actually fired when compiling the miscompilation sample provided as a part of the bug tracker report.
It is of course possible that the ``fixing'' patch does not entirely fix the miscompilation, and/or introduces new bugs in the compiler; sometimes bug reports are reopened for just this reason (see Clang/LLVM bug \#21903\footnote{\url{https://bugs.llvm.org/show_bug.cgi?id=21903}} 
for an example). We are reasonably confident that the bugs used in our study do not fall into this category: their fixes, accepted by the open-source community, have stood the test of time.

For some patches it was tractable to determine precise conditions under which a compilation fault would have occurred in the buggy compiler.  However, in other cases it was difficult or impossible to determine such precise conditions, either because the code associated with the patch was complex enough that determining the exact conditions under which the patch exhibits erroneous behaviour would require highly-specialized knowledge, or because the occurrence of a fault could not be verified for sure based on available compile-time information only.  In these cases we instead settled for \textit{over-approximating conditions},
designed to certainly issue warnings when the fault is triggered, but possibly issuing false positives, \ie warning that the fault is triggered when it is not.  In such cases we worked hard to make the over-approximating conditions precise to the best of our abilities. As an example, Clang/LLVM bug \#21242 (reported by the Alive tool) affects an IR code transformation: a multiplication operation between a variable and a power of two, $x*2^n$, is transformed into the shift left operation $x<<n$. When $n=31$, the transformation is faulty for 32-bit signed integers in case when $x=1$, because the overflow semantics of the first operation is not correctly preserved within the second one. As the value $x$ will hold at runtime is unknown at compile time (and may be different each time the compiled code is executed), we must conservatively issue a fault warning regardless of the value of $x$.
%



\subsection{Stage 2: Syntactic Binary Analysis}
\label{sec:stage-two}

The first stage of our approach provides us with better insight into how the application's compilation process is affected by the bug, including information on certain (in the case of precise conditions) and potential (in the case of over-approximating conditions) compilation faults. The second stage is focused on the result of a possible faulty compilation: it performs a syntactic comparison of the application binaries produced by the buggy and fixed compilers, to understand how much the possible compilation fault made them differ syntactically.

As a preliminary check, we perform a bitwise comparison of the monolithic application binaries generated by the buggy and fixed compilers. If these monolithic binaries are different, we disassemble them and textually compare the two versions of the assembly code, in order to estimate how many of the defined assembly functions exhibit significant syntactic differences induced by the compilation fault. In particular, our function-by-function comparison includes heuristics (see \S\ref{sec:exp-infrastructure}) designed to ignore irrelevant differences, such as specific addresses.

Observe that for Stage 2 to be meaningful, we need to make sure that the detected differences are only caused by the code differences in the two compilers (\ie by the fixing patch). In practice, this may not always be the case, as the compilation process of some application may not be reproducible.
In such cases, we could mistakenly infer that the differences in the
two binaries are caused by a compilation fault. We discuss in
\S\ref{sec:sample-apps} how we have practically ensured that the
applications compiled in our study follow a reproducible build
process.

If Stage~1 determines that no fault was triggered during application compilation, the binaries produced by the buggy and fixed compilers are expected to be identical.  In such cases, we still perform Stage~2 to sanity-check our approach: if Stage~1 reports no triggered faults but Stage~2 reports differences in generated binaries, it might indicate that something is wrong either with the way the warning-laden compiler has been hand-crafted or with the reproducibility of the application build process.  In practice, this careful approach led to us to detect that the binaries produced for some applications made use of the revision number of the compiler used to build them. As the buggy and fixed compilers correspond to different revisions, the binaries that they produced for these applications were always different, even when no compilation fault occurred. We solved this problem by removing any mention of the revision number within the compilers that we used.

Finally, observe that when Stage~1 of our approach detects that a fault was triggered during application compilation,  the binaries produced by the buggy and fixed compilers might yet be identical. This can be due to a false alarm at Stage~1 resulting from over-approximating conditions, but also to
cases where an actual fault is masked at a later stage of compilation.

\subsection{Stage~3: Dynamic Binary Analysis}
\label{sec:stage-three}

Even if Stage~2 discovers that the binaries produced by the buggy and fixed compilers differ, this does not guarantee that the possible compilation fault detected at Stage~1 propagates to a miscompilation failure.
Indeed, Stage~2 only evaluates how much the binaries produced by the
buggy and fixed compilers are \textit{syntactically} different, which
does not prevent them from being \textit{semantically} equivalent.  In
this case, the differences observed in the binary compiled by the
buggy compiler are not the witnesses of a miscompilation, as they
cannot trigger any incorrect application behaviour.


The purpose of the third stage of our approach is to understand
whether the compilation fault propagates to a miscompilation
failure. To study this, we run the default regression test suite of
the application once with the binary produced by the buggy compiler,
and once with the binary produced by the fixed compiler.  If the first
binary fails on some tests while the second one does not, it
means---modulo some possibly flaky tests~\cite{covrig,flaky:fse14} or
some undesirable non-determinism in the application---that the fault
can make the application misbehave at runtime and thus propagates
to a failure.  If the test results are the same for the two binaries,
it is unknown whether this is because they are semantically equivalent
or because the test suite is not comprehensive enough to expose the
miscompilation failure.

We note that the effectiveness of Stage~3 is notably impacted by the
quality of the application's test suite, and in particular whether the
test suite is deemed thorough enough to act as a proxy for typical
real-world usage. Another important point is that if the test suite of
a particular application is regularly used to exercise a compiler,
leading to miscompilations being detected and fixed as they arise, the
application's test suite might subsequently be considered as immune to
all the bugs affecting the versions of the compiler that it has
exercised.
We discuss in \S\ref{sec:threats} this risk and how we have selected
applications in a manner that minimises it.

To mitigate the limitations associated with using test suites only, if
no differences are observed when running the test suites for a
compiler bug, we may sample a set of the syntactic differences spotted at
Stage~2 in the assembly code of the applications impacted by the
bug. Each syntactic difference in the sample is then manually
inspected with an eye to crafting inputs that would make the two binaries diverge at
runtime because of this difference. Depending on the complexity of the
application, these inputs can either be global inputs to the
application leading to a divergence in the application's behaviour, or
inputs local to the function affected by the syntactic difference that
trigger a semantic divergence at this local level (such that we do not
know---due to the application's complexity---whether there is an input
to the application that would lead to invocation of the function with
the local input). Once all the sampled syntactic differences have been
investigated, we try to generalise the gained knowledge, complemented
by our understanding of the bug and fixing patch's details on the bug
tracker, into a rule of thumb assessing the impact that binary differences induced by the compiler
bug can have over the semantics of an application.


%% file: 4infrastructure.tex


We now describe how we chose the bugs  (\S\ref{sec:sample-bugs}) and applications (\S\ref{sec:sample-apps}) considered in the study, and discuss the technical aspects of the chosen experimental process (\S\ref{sec:exp-infrastructure}) .

\subsection{Sampling Compiler Bugs to Investigate}
\label{sec:sample-bugs}

Due to the steep learning curve associated with gaining expertise in a production compiler framework, and the intensive manual effort required to prepare warning-laden compilers for each bug we consider, we decided to restrict attention to bugs reported in a single compiler framework.  Most publicly-reported compiler fuzzing efforts have been applied to Clang/LLVM and GCC.  Either would have been suitable; we chose to focus on Clang/LLVM as this enabled an interesting comparison between bugs found by fuzzers and bugs found as a by-product of formal verification (the Alive tool is not compatible with GCC).

A total of 1,033 Clang/LLVM bugs are listed within the scoreboards of the four C fuzzers and Alive at time of writing.
Relevant properties of these bugs are summarised in Table~\ref{tab:sampling}.

Our study requires a fixing patch for each bug, and our aim is to study miscompilations.  We thus discarded the 799 bugs that remain unfixed or are not miscompilation bugs.

{\scriptsize
\begin{table}[t]
 \centering
 \caption{ Tool-reported Clang/LLVM bugs that we study.}
     \SPACEHACK{-2mm}
 \label{tab:sampling}
\begin{tabular}{|c|c|ccc|}
\hline
\multirow{3}{*}{\textbf{Tool family}} & \multirow{3}{*}{\textbf{Tool type}} &\multicolumn{3} {c|}{\textbf{Number of Clang/LLVM bugs}} \\
 & & Bugs & Fixed & Final  \\
 & & reported & miscompilations & sample  \\
\hline
\hline
Csmith & \multirow{4}{*}{fuzzer} & 164 & 52  & 10   \\
EMI   &  & 783 & 163  &  10    \\
Orange  & & 12  & 7 &  5  \\
yarpgen  & & 66  & 4  &  2    \\
\hline
Alive & formal verifier & 8  & 8  &  8    \\
\hline
\hline
\multicolumn{2} {|c|}{\textbf{TOTAL}} & \textbf{1033}  &  \textbf{234}   &   \textbf{\numFuzzerAndAliveBugs}     \\
\hline
\end{tabular}
\end{table}
}

We then removed any of the remaining bugs for which the affected Clang/LLVM versions are too old to be built from source and to compile packages within a Debian 9 installation (with reasonable effort). In practice, this means that we exclude all the bugs affecting Clang/LLVM versions older than 3.1 (which was released more than seven years ago).

Because the aim of our study is to assess the extent to which miscompilation bugs have high practical impact, we pruned those bugs that only trigger when non-standard compilation settings are used, e.g.\ to target old or uncommon architectures or to enable optimisation levels higher than default for Debian packages (-O2).

Finally, we randomly selected 10 bugs (if available\footnote{e.g. there are only 7 fixed miscompilation bugs in Clang/LLVM that were found by the Orange tools, out of which two were excluded because they could not be triggered with standard compilation settings.}) to study per tool, among those bugs for which we were able to isolate an independent fixing patch and write a corresponding warning-laden compiler. This led to a final sample of \numFuzzerAndAliveBugs bugs to analyse, as shown in Table \ref{tab:sampling}. We completed this sample by adding a set of 10 miscompilation bugs reported directly by end user of Clang/LLVM. These bugs were selected by searching the Clang/LLVM bug tracker for a set of 20 suitable miscompilation bugs not reported by the fuzzers or Alive authors. Then, we picked the first 10 of these bugs for which we could isolate an independent fixing patch and write a corresponding warning-laden compiler.  

Overall, the studied sample of bugs covers 12\% of the fixed miscompilation bugs found in Clang/LLVM by the four considered fuzzers, all the bugs found by the Alive tool and 10 bugs reported by end-users, which we regard as a significant but tractable subset.  While studying more bugs would have made this study even stronger, we were limited principally by the human effort required per bug (and to some degree also by machine time).

\subsection{Sampling Applications to Compile}
\label{sec:sample-apps}

As the set of applications to be compiled, we consider the source packages developed for Debian (version 9), the most stable release at the time when we conducted our study. More than 30,000 packages are available.  Experimenting with all packages was infeasible in terms of the machine resources available to us.  We now describe the process that we followed to select a feasible number of packages.

To apply the methodology of \S\ref{sec:methodology} to C/C++ compiler
bugs in an automated fashion, we required packages that: build in a
reproducible fashion (a requirement of Stage 2,
see \S\ref{sec:stage-two}); are not used for exercising Clang/LLVM on
a regular basis and come with sufficiently thorough test suites (both
requirements of Stage 3, see \S\ref{sec:stage-three}) which can be
executed in a standard fashion (required for automation); and contain
more than 1K lines of C/C++ code (a limit that we imposed to filter out
uninteresting applications).

For build reproducibility, the Debian Reproducible Builds initiative
%
lists the approximately 23,000 packages for which builds are believed to be deterministic.
Regarding availability of test suites, about 5,000 packages support the Debian Autopkgtest command as a unified interface for running tests.  We thus restricted attention to the approximately 4,000 packages with reproducible builds and Debian Autopkgtest-compatible test suites.

We then filtered out all the packages that contain fewer than 1K lines of C/C++ code, using the \textit{cloc}\footnote{\url{http://cloc.sourceforge.net}} tool to count lines of code.  This was important in order to filter out trivial applications or applications that are mostly written in another programming language but that include a small C/C++ component.

To sanity-check reproducibility of package builds, we built each selected package twice with the same compiler (an instance of version 3.6 of Clang/LLVM) and verified that bitwise identical binaries are indeed produced.


To make the running time of our analyses more acceptable (it could take more than one week per compiler bug considering all the remaining packages), we sampled half of these packages to end up with a set of \numPackages source packages, consisting of a total of 8,283,744 lines in C source files, 782,703 lines in C++ source files and 1,250,948 lines in C/C++ header files.
In more detail, 179 packages consist of between 1K to 10K lines of code in C/C++ source or header files, 108 packages between 10K to 100K, 21 packages between 100K to 1M and one package more than 1M. The sampling was performed first by identifying a set of about 50 popular and 50 limited-audience representatives\footnote{We used \url{https://popcon.debian.org} to estimate popularity.} of a wide variety of application types, such as system utilities (\grep), web servers (\apache), scientific software (\textit{Symmetrica}), network protocols (\samba) and printer drivers (\textit{Epson-inkjet-printer-escpr}). We randomly selected other packages to complete the sampling.

During package sampling, we also checked that none of the selected
packages are part of the extended test suite used by the Clang/LLVM
developers to exercise regularly the compiler.  We discuss our
rationale for doing this in \S\ref{sec:threats}.


In order to evaluate test suite thoroughness for the packages to be
chosen, we use
the \textit{gcov}\footnote{\url{https://gcc.gnu.org/onlinedocs/gcc/Gcov.html}}
tool to gather statement coverage information for the package test
suites.  However, integrating gcov within the package build
infrastructure presented issues, \eg gcov notes files were not
always generated during compilation, sometimes coverage data were not
generated during test suite runs, and sometimes both were generated
but were deemed not to match.  Because of these issues, gcov was able to
produce reliable coverage data only for a sample of 39 packages.  Across
these packages, we found their test suites to achieve a median of 47\%
and a mean of 46\% statement coverage, with lowest and highest
coverage rates of 2\% and 95\%, respectively.  About half of the
packages---18/39---had test suites achieving at least 50\% statement
coverage. While statement coverage is a limited metric, these coverage
rates were somewhat higher than our team had predicted.

\subsection{Experimental Process}
\label{sec:exp-infrastructure}

We now describe the technical aspects of the process that we followed to measure the impact of the \numAllBugs sampled compiler bugs over the \numPackages sampled Debian packages, using the three-stage methodology described in \S\ref{sec:methodology}.

Experiments were performed over six local servers and ten private cloud instances, except for the test suite runs, which were conducted within instances set up using Amazon Web Services (AWS).\footnote{\url{https://aws.amazon.com}} 
Each bug was analysed in one Debian 9 virtual machine installed either on the local servers or on the remote private cloud instances, while test suite runs necessary for such an analysis were performed over two Debian 9 AWS machines, specifically created for each bug.

As a preliminary step, we prepared the list of the \numPackages packages to analyse, together with exact package names and versions, in a JSON format that allows Simple Build to automatically download the corresponding source packages from the Debian website.

For each bug, the analysis starts by instructing Simple Build to install a fresh local Debian 9 build environment in a chroot jail. The sources of the warning-laden, buggy and fixed compilers are then compiled and installed in this build environment. Finally, we iterate over the packages defined in the JSON file and perform the three stages detailed in our methodology for each of them.

Stage 1 (\S\ref{sec:stage-one}) is performed by setting the warning-laden compiler as the default compiler in the build environment and asking Simple Build to build the package. The resulting build logs are then searched (using \textit{grep}) for the warning messages. In some cases, Simple Build may fail because the package build process is not compatible with the Clang/LLVM versions affected by the bug. These cases are simply logged and Stages 2 and 3 are not carried out.

Stage 2 (\S\ref{sec:stage-two}) is performed by setting successively the buggy and fixed compiler as the default compiler in the build environment and asking Simple Build to build each time the package. The two resulting binaries are then compared bitwise using \textit{diff}. 
If the two resulting binaries are different, we disassemble their code sections (using \textit{objdump}). The functions defined in the two disassembled binaries are then textually compared opcode-by-opcode and we count the ones that differ.
By comparing only the opcodes we aim to achieve a good trade-off between false positives and false negatives. To avoid false positives, we do not consider the differences in the operands, since these differences, namely addresses, registers and immediate values, may not affect the semantics of the functions. However, aggressively leaving out all the operands may also lead us to overlook the compiler bugs that only affect the operands in the binaries produced by the buggy compiler, potentially leading to false negatives. Yet, none of the bugs considered in this study is likely to produce such false negatives (see \S\ref{sec:threats} for a more detailed discussion).


Stage 3 (\S\ref{sec:stage-three}) is performed by asking Autopkgtest to execute the package test suite over the two binaries produced at Stage 2, if they were different. The two test runs are carried out within the two isolated AWS test environments. 
The two resulting test logs are then hand-checked and compared. When a
difference is spotted, the test runs are repeated several times in
fresh test environments to make sure that the results are reproducible
and not polluted by flaky tests or non-determinism. For some packages,
the testing infrastructure may not be reliable---we log the cases
where the infrastructure crashes, making it impossible to run the
tests. In case no divergences are spotted in the test results, a
sample of the differently-compiled functions are randomly selected for
some bugs and manually inspected. For each function, we
use \textit{gdbgui}\footnote{\url{https://gdbgui.com/docs.html}} to recover the part of the source code that is
compiled differently.  If this code can be easily reached
from the executable's main function and if the program has simple
inputs and outputs, we manually try to construct inputs to the program
that result in runtime divergences. If the code cannot be easily
reached from the main function or if testing the executable as a whole
is too complex, we isolate and compile a driver for the C/C++ function
that contains the identified code and try to trigger a divergence
similarly.

We estimate the total machine time spent to run all the experiments to be around 5 months.


%% file: 5results.tex

We now analyse the results obtained for every bug and package pair considered in our study.
We discuss the results for the fuzzer-found bugs in detail, covering the three stages of our approach in \S\ref{sec:compile-time-analysis}---\S\ref{sec:test-suite-execution}, and turn to a comparison with bugs found from other sources---user-reported bugs and Alive bugs---in \S\ref{sec:comparison-with-other-bug-sources}.
We also investigate whether there is a correlation between bug severity and impact in \S\ref{sec:severity-and-impact}.  We wrap up with a discussion of our overall experimental findings in \S\ref{sec:discussion}.

{\scriptsize
\begin{table*}[h]
    \centering
    \caption{Impact analysis for \numFuzzerBugs Clang/LLVM bugs found by the 4 fuzzers families over \numPackages Debian packages.}
    \SPACEHACK{3mm}
    \label{tab:impactLLVMfuzzers}
\begin{tabular}{|cc||c||ccc|cc|cc|}
\hline
\multicolumn{2} {|c||}{\textbf{BUG}} & \textbf{PACKAGES} & \multicolumn{3} {c|} {\textbf{1) BUGGY LLVM CODE}} &  \multicolumn{2} {c|}{\textbf{2) BINARY DIFFS}} & \multicolumn{2} {c|} {\textbf{3) RUNTIME DIFFS}} \\
\textit{id} & \textit{severity} & \textit{successful builds} & \textit{reached} &  {\textit{triggered}}  &  {\textit{(precise)}} & \textit{packages} & \textit{functions} & \textit{test diffs} & \textit{manual}  \\
\hline
\hline
\multicolumn{10}{|l|}{\textbf{Csmith} (10)}   \\
\hline
 11964 & enhancement     & 307 & 306 &   2 & (no)  &   2 & < 0.1\% [52]     & 0 & \textit{low}  \\
 11977 & normal          & 307 & 301 & 118 & (no)  &  20 & < 0.1\% [35]     & 0 & - \\
 12189 & enhancement     & 307 & 297 & 291 & (no)  &  46 & < 0.1\% [177]    & 0 & - \\
 12885 & enhancement     & 304 & 284 &   1 & (no)  &   0 &      -      & - &  - \\
 12899 & enhancement     & 306 & 143 &   6 & (no)  &   0 &        -     &  - & - \\
 12901 & enhancement     & 306 & 291 & 286 & (no)  &  36 & < 0.1\% [50]     & 0 & - \\
 13326 & enhancement     & 304 & 125 & 125 & (no)  &   0 &     -        & - & - \\
 17179 & normal          & 305 & 245 &   3 & (no)  &   2 & < 0.1\% [7]      & 0 & - \\
 17473 & release blocker & 308 & 285 &  16 & (no)  &  10 & < 0.1\% [16]     & 0 &  -\\
 27392 & normal          & 308 & 205 & 205 & (yes) & 202 & 2.5\% [4997] & 0 & \textit{very low} \\
\hline
\multicolumn{2} {|c||}{\multirow{ 2}{*}{\textbf{TOTAL}}}  & {\textbf{3062}}    & {\textbf{2482}} & {\textbf{1043}}  & & {\textbf{318}} & {\textbf{0.4\% [{\textbf{5334}}]}}& {\textbf{0}} &    \\
  &  & {\textbf{100\%}}   & {\textbf{81\%}} & {\textbf{34\%}}  & &  {\textbf{10\%}}   &  &  {\textbf{0\%}}     &   \\
\hline
\hline
\multicolumn{10}{|l|}{\textbf{EMI} (10)}   \\
\hline
 24516 & normal & 307 & 130 &   0 & (yes) &   0 &       -    & - & -  \\
 25900 & normal & 307 & 221 &   5 & (no)  &   0 &        -   & - & - \\
 26266 & normal & 308 & 302 & 195 & (no)  &   0 &      -      & - & - \\
 26323 & normal & 305 & 281 &  32 & (no)  &  12 & < 0.1\% [18]    & 0 & \textit{very low} \\
 26734 & normal & 308 & 175 &   5 & (no)  &   0 &      -      & - & - \\
 27968 & normal & 308 & 122 &   0 & (yes) &   0 &      -     & - & - \\
 28610 & normal & 306 & 300 & 295 & (no)  &   9 & < 0.1\% [15]    & 0 & - \\
 29031 & normal & 307 & 297 & 215 & (no)  & 127 & 0.3\% [639] & 1 & \textit{low}  \\
 30841 & normal & 308 & 306 & 191 & (no)  &   0 &      -     & - & - \\
 30935 & normal & 308 & 287 &  10 & (no)  &   3 & < 0.1\% [3]     & 0 & \textit{low} \\
\hline
\multicolumn{2} {|c||}{\multirow{ 2}{*}{\textbf{TOTAL}}}   & {\textbf{3076}}    & \textbf{2424}  & {\textbf{948}}  & & \textbf{151} & \textbf{< 0.1\%} {\textbf{[675]}} & 1 &    \\
  & & {\textbf{100\%}}   & \textbf{79\%}  & {\textbf{31\%}}  & &  \textbf{5\%} &  & \textbf{< 0.1\%}  &   \\
\hline
\hline
\multicolumn{10}{|l|}{\textbf{Orange} (5)}  \\
\hline
 15940 & normal & 307 & 158 & 19 & (no)  & 0 &     -    & - & - \\
 15959 & normal & 307 & 108 &  9 & (no)  & 8 & 0\% [14] & 0 & - \\
 19636 & normal & 307 &   7 &  7 & (no)  & 0 &     -   & - & - \\
 26407 & normal & 308 &   4 &  0 & (yes) & 0 &     -  & - & - \\
 28504 & normal & 306 &  16 &  0 & (no)  & 0 &     -    & - & - \\
\hline
\multicolumn{2} {|c||}{\multirow{ 2}{*}{\textbf{TOTAL}}}  & {\textbf{1535}}    & {\textbf{293}}  & {\textbf{35}} &   & {\textbf{8}}  & {\textbf{< 0.1\% [14]}}   & {\textbf{0}}  &    \\
  &  & {\textbf{100\%}}   & {\textbf{19\%}} & {\textbf{2\%}} & & {\textbf{1\%}} && {\textbf{0\%}} &  \\
\hline
\hline
\multicolumn{10}{|l|}{\textbf{yarpgen} (2)}  \\
\hline
 32830 & enhancement & 308 & 301 &   0 & (yes) & 0 & - & - & - \\
 34381 & enhancement & 307 & 307 & 257 &  (no) & 0 & - & - & - \\
\hline
\multicolumn{2} {|c||}{\multirow{ 2}{*}{\textbf{TOTAL}}}  & {\textbf{615}}    & {\textbf{608}}  & {\textbf{257}} & & {\textbf{0}}&   &   &  \\
  &  & {\textbf{100\%}}   &  {\textbf{99\%}}  & {\textbf{42\%}} &  {\textbf{}}  &  {\textbf{0\%}} & {\textbf{}}  &  {\textbf{}} &  {\textbf{}}  \\
\hline

\hline
\hline
\multicolumn{2} {|c||}{\multirow{ 2}{*}{\textbf{ALL}}}  & {\textbf{8284}}    & {\textbf{5804}}  & {\textbf{2283}} & & {\textbf{477}}&  {\textbf{0.2\% [6023]}}  &  {\textbf{1}}  &  \\
&  & {\textbf{100\%}}   &  {\textbf{70\%}}  & {\textbf{28\%}} &  {\textbf{}}  &  {\textbf{6\%}} & {\textbf{}}  &  {\textbf{< 0.1\%}} &  {\textbf{}}  \\
\hline
\end{tabular}
\end{table*}
}
{\scriptsize
\begin{table*}[h]
    \centering
    \caption{Data of Table \ref{tab:impactLLVMfuzzers}, aggregated by fuzzer and by compiler bug.}
        \SPACEHACK{-3mm}
    \label{tab:impactLLVMfuzzersbugs}
\begin{tabular}{|c||c||ccc|c|c|}
\hline
{\multirow{ 2}{*}{\textbf{TOOL}}} &  \multirow{ 2}{*}{\textbf{BUGS}}  & \multicolumn{3} {c|} {\textbf{1) BUGGY LLVM CODE}} & {\textbf{2) BINARY DIFFS}} &   {\textbf{3) RUNTIME DIFFS}} \\
    & & \textit{reached} & \textit{triggered} & \textit{(precise)}  & \textit{packages} &  \textit{test diffs}   \\
\hline
 Csmith & 10 & 10 & 10 & (1) & 7 & 0  \\
 EMI & 10 & 10 & 8 & (0) & 4 & 1  \\
 Orange & 5 & 5 & 3 & (0) & 1 & 0  \\
 yarpgen & 2 & 2 & 1 & (0) & 0 & -   \\
\hline
 \textbf{TOTAL} & {\textbf{27}}   & {\textbf{27}} & {\textbf{22}} & ({\textbf{1}}) & {\textbf{12}} & {\textbf{1}}       \\
\hline
\end{tabular}
\end{table*}
}

\subsection{Stage 1: Compile-time Analysis}\label{sec:compile-time-analysis}

Experimental results for the fuzzer-found bugs are presented in
Table~\ref{tab:impactLLVMfuzzers}, with
Table~\ref{tab:impactLLVMfuzzersbugs} providing a more condensed view
with results aggregated by fuzzer and by compiler bug.

\myparagraph{Package Build Failures.} Fewer than 1\% of all package builds
failed. All but one of the analysed compiler versions failed to build at
least one package; the maximum number of packages that a specific compiler version failed to build was five. Across all compiler versions,
build failures were associated with \numFailedBuildPackages particular packages.  The package with the
highest failure rate is the \textit{Velvet} bioinformatics tool, for
which \percentVelvetFailure of the builds failed. Manual inspection of build failure logs
shows front-end compilation errors, e.g.\ relating to duplicate
declarations and missing header files.

\myparagraph{Reachability of Fixing Patch.} For each bug, at least one package
caused the associated fixing patch to be reached during compilation; \ie all the fuzzer-found bugs we studied were related to code that could be reached during compilation of standard packages.
Note also that for each of our \numPackages packages, the compilation process reached the fixing patch of at least one of the bugs that we study.

For \numPatchReachMoreThanHalfFuzzerBugs/\numFuzzerBugs bugs, the proportion of packages for which
the patch was reached is above 50\% and it remains high for most of
the other bugs. The highest proportion is attained with Yarpgen
bug \#34381, whose patch is reached for all 307 packages that built successfully. This bug
affects the generation of x86 binary code for additions. The minimal
proportion is attained with Orange bug \#26407, whose patch is reached
for fewer than 2\% of the packages. This bug affects the remainder
operation for unsigned numbers when it involves a power of two. In
general, the proportion of packages where the patch is reached appears
to be much lower for the bugs discovered by Orange than by the other
tools. A likely explanation is that Orange focuses on bugs affecting potentially complex arithmetic, which does not appear so commonly in
real-world code.



\myparagraph{Fault Triggering.} We were able to come up with precise
fault-triggering conditions for only 19\% of the investigated
bugs. The main difficulty in writing such conditions was that the
fixing patch for many bugs makes it difficult to identify the precise
situations where the buggy version of the code would fail. Indeed, for
many patches, it is highly complex to determine how the local changes
made by the patch precisely impact the global behaviour of the
compiler.

Due to our best efforts to make the imprecise patches as precise as
possible, only \percentTriggerByReachFuzzerBugs of the builds where a
fixing patch is reached lead to the fault conditions triggering.  In
total, \numMaybeTriggeredFuzzerBugs/\numFuzzerBugs compiler bugs
and \percentTriggerByBuildFuzzerBugs of the package builds generated
potential faults. This last number falls
to \percentTriggerByBuildPreciseFuzzerBugs when restricted to the
cases where the patch is precise and thus the fault is certain.

\subsection{Stage 2: Syntactic Binary Analysis}\label{sec:binary-comparison}

While Stage~1 returned a \percentTriggerByBuildFuzzerBugs possible
fault rate, only \percentBinDiffByBuildFuzzerBugs of the package
builds actually led to different binaries being generated by the buggy
and fixed compiler. This difference is mainly due to false alarms
issued by our fault detectors in the case of imprecise
conditions. However, manual inspection revealed that in some cases
this can also be caused by the particularities of the package build
process. For example, in the context of Csmith bug \#27392, there are
three packages,
namely \textit{libffi-platypus-perl}, \textit{snowball}
and \textit{viennacl}
where a fault is found to be triggered at Stage~1, but identical binaries
are produced at Stage~2. This is due to the fact that the binary for which
the fault was triggered
is not included as a part of the final packaged binary returned by the build
process.


Regarding the aggregated numbers obtained for the
Csmith bugs (see Table~\ref{tab:impactLLVMfuzzers}),
the buggy compiler code is reached for \percentReachCsmith of the builds and the fault conditions are triggered during \percentTriggerCsmith of the builds. Similar numbers
are obtained for the EMI bugs: \percentReachEMI and \percentTriggerEMI respectively.
However, the possible fault rate in terms of packages at Stage~2 decreases almost
twice as fast for EMI (\percentBinDiffEMI) as for
Csmith (\percentBinDiffCsmith), resulting in \numBinaryDiffsEMI
out of \numEMIBugs EMI bugs causing binary differences against \numBinaryDiffsCsmith out of \numCsmithBugs for
Csmith. While this trend should be confirmed using a larger number of
bugs, a possible explanation is that Csmith was the first tool used to
fuzz Clang/LLVM, so it had the privilege to collect some
``low-hanging fruit'', \ie those compiler bugs that impact binaries more often.

In total, \numBinaryDiffsFuzzerBugs/\numFuzzerBugs fuzzer-found bugs (44\%) lead to differences in the generated monolithic binaries for at least one package. The compilers affected by bugs Csmith \#27392 and EMI \#29031 caused 53\% of 
the package builds leading to such differences. These two bugs both affect an optimisation pass, respectively loop unrolling and hoisting of load/store instructions to a dominating code block.

Among the 12 bugs that lead to different binaries, 10 affect only very few (at most 177) of all the assembly functions generated when compiling our list of packages, accounting for less than 1\textperthousand\ of them. Note that three packages, \modgearman, \phyml and \samba, are not taken into account because the build process removes some information, e.g. debug symbols, required by our analysis.
Bug Csmith \#27392 hits a record of 2.5\%, while  EMI \#29031 affects 0.3\% of the generated assembly functions. Notice that the total number of assembly functions in the binaries of our list of packages may vary with the version of Clang/LLVM used to compile them. Indeed, as each version may come with a different implementation of the optimisation passes, the compiler may decide on a different set of function calls to inline. We counted the total number of assembly functions in the binaries produced by the buggy compilers for bugs Csmith \#11977,  Alive \#20189 and Csmith \#27392, representing respectively an early, medium and late point in the timeline of the compiler versions used in our study. The average, around 202K, with a standard deviation of 5\%, provides an approximate value that we used to compute the aforementioned percentages.

Finally, for one package and bug pair, it happened that the build process failed using the buggy compiler, while it was successful using the fixed compiler. A manual inspection revealed that the build process included running the package test suite after compilation, and that a test case was failing when the package had been compiled by the buggy compiler. This test failure was caused by a miscompilation induced by the investigated compiler bug, as detailed in the next section.

\subsection{Stage 3: Dynamic Binary Analysis}\label{sec:test-suite-execution}

\myparagraph{Failed Test Suite Runs.} About \percentTSFailFuzzers of the test suite
runs could not be carried out because the underlying testing
infrastructure was not reliable and crashed. The most common reasons
for crashes were missing files or dependencies and
failures to compile the testing infrastructure itself. This does not necessarily indicate that those test suites do not work, but rather that they did not work seamlessly with our automated approach, and that it was not feasible, time-wise, to look at them manually.

\myparagraph{Differences in Test Results.} Across all bugs and packages,
we have only observed a single test failure: when compiling the \textit{zsh} package with the buggy compiler affected by bug EMI \#29031 (the technical details of this miscompilation are available in the appendix \S\ref{sec:miscomp-details});
\ie it would appear that the impact of these compiler bugs is not severe enough to cause frequent failures in the current regression test suites of these packages.

%
%
%
%

\myparagraph{Additional Experiments.} Given the rather unexpected
result that only one compiler bug was able to trigger a single test failure within
our \numPackages packages, we repeated our analysis over
\sqlite,\footnote{\url{https://www.sqlite.org}} a large, complex and
widely-used application reputed for the thoroughness of its testing
process (its free default test suite achieves more than 98\% coverage
of the application's 151K statements). 

This additional analysis did not reveal any difference in the test
results, except for CSmith
bug \#13326,\footnote{\url{https://bugs.llvm.org/show_bug.cgi?id=13326}}
where the binary produced by the buggy compiler fails on test
case \textit{incrblob-2.0.4}, while the binary generated by the fixed
compiler does not. Interestingly, the miscompilation
responsible for this failure does not affect the \sqlite code itself,
but the code of the test bed responsible for conducting
test \textit{incrblob-2.0.4}. Again, the interested readers can find a technical description of this miscompilation in the appendix \S\ref{sec:miscomp-details}.

It is interesting to remark that while \sqlite is part of the Clang/LLVM test suite (so that it was excluded from our selected list of packages), this has not enabled detecting and fixing bug \#13326 before a fuzzer did. This is likely due to the fact that once the Clang/LLVM test suite has compiled \sqlite, it only runs it over two small tests that barely exercise its features. In contrast, the \sqlite test suite that we have used to expose the miscompilation in this experiment contains more than one thousand high-coverage tests. Finally, notice that test \textit{incrblob-2.0.4} is
also miscompiled and fails for the binaries produced by both the buggy
and fixed compilers of Clang/LLVM bugs \#11964, \#11977,
\#12189 and \#12885, as these compiler versions are all within the lifespan
of bug \#13326.



\myparagraph{Manual Inspection.} Given the near-absence of any differences
in the test results obtained for all the investigated compiler bugs,
we have manually inspected a sample of the syntactic binary
differences detected at Stage~2. This inspection
reveals that a significant number of the studied binary differences have no impact over the semantics of the
application, because compiler developers tend to be rather
conservative when fixing bugs, \eg deactivating many legitimate
code transformations as a side-effect of the fix. In case the binary-level difference is seen to possibly impact the application's semantics, the manual
inspection suggests that very specific circumstances are typically
required to happen at runtime for the miscompilation to trigger a
corruption of the program's internal state and for this corruption to
propagate to the program's visible outputs. In many cases, these
circumstances are very unlikely or simply impossible. For example, bug
\#11964 affects only 52 functions and after manually analysing all of
them, we think none can trigger a runtime failure. For interested readers, we provide a detailed discussion of the most salient
technical results of our manual inspection in the appendix \S\ref{sec:manual-anal}.

\subsection{Comparison with Other Bug Sources}\label{sec:comparison-with-other-bug-sources}

{\scriptsize
\begin{table*}[h]
    \centering
    \caption{Impact analysis for the 18 Clang-LLVM bugs found by Alive or end users over \numPackages Debian packages.}
        \SPACEHACK{-3mm}
    \label{tab:impactLLVMalive}
\begin{tabular}{|cc||c||ccc|cc|cc|}
\hline
\multicolumn{2} {|c||}{\textbf{BUG}} & \textbf{PACKAGES} & \multicolumn{3} {c|} {\textbf{1) BUGGY LLVM CODE}} & \multicolumn{2} {c|} {\textbf{2) BINARY DIFFS}} & \multicolumn{2} {c|} {\textbf{3) RUNTIME DIFFS}} \\
\textit{id} & \textit{severity} & \textit{successful builds} & \textit{reached} &  {\textit{triggered}}  &  {\textit{(precise)}} & \textit{packages} & \textit{functions}  & \textit{test diffs} & \textit{manual} \\
\hline
\hline
\multicolumn{10}{|l|}{\textbf{Alive} (8)}  \\
\hline
 20186 & normal & 309 &  34 &   0 & (yes) &   0 &  -          & - &  \\
 20189 & normal & 309 & 266 & 176 & (no)  & 122 &  1\% [2094] & 0 &  \\
 21242 & normal & 309 & 253 & 151 & (no)  &  50 &  0\% [130]  & 0 &  \\
 21243 & normal & 309 &  56 &   0 & (yes) &   0 &  -          & - & \\
 21245 & normal & 309 & 274 &   0 & (yes) &   0 &  -          & - & \\
 21255 & normal & 309 &   9 &   0 & (yes) &   0 &  -          & -&  \\
 21256 & normal & 309 & 167 &   0 & (yes) &   0 &  -          & -&  \\
 21274 & normal & 309 &   0 &   0 & (yes) &   0 &  -          & - & \\
\hline
\multicolumn{2} {|c||}{\multirow{ 2}{*}{\textbf{TOTAL}}}   &{\textbf{2472}}      & {\textbf{1059}}  & {\textbf{327}}  & & {\textbf{172}} &  {\textbf{0.6\% [2224]}}   & {\textbf{0}}  &    \\
 & &{\textbf{100\%}}     &  {\textbf{43\%}}    & {\textbf{13\%}} & & {\textbf{7\%}}   & & {\textbf{0\%}}  &      \\
\hline
\hline
\multicolumn{10}{|l|}{\textbf{User-reported} (10)}  \\
\hline
 13547 & release blocker & 306 & 300 & 278 & (no)  &  0 & -         & - & \\
 15674 & release blocker & 307 & 301 &   0 & (yes) &  0 & -         & -  & \\
 17103 & release blocker & 305 & 305 &   0 & (no)  &  0 & -         & - & \\
 24187 & normal          & 309 &   0 &   0 & (yes) &  0 & -         & - & \\
 26711 & normal          & 308 &   0 &   0 & (yes) &  0 & -         & - & \\
 27575 & normal          & 308 & 133 &  44 & (no)  &  0 & -         & - &  \\
 27903 & normal          & 308 & 286 & 231 & (no)  & 52 & < 0.1\% [169] & 1 & \textit{low} \\
 31808 & normal          & 308 & 229 &   0 & (yes) &  0 & -         & - & \\
 33706 & normal          & 308 & 259 &  40 & (no)  &  4 & < 0.1\% [9]   & 0 & \textit{low}  \\
 37119 & normal          & 303 & 177 &   0 & (no)  &  0 & -         & -  & \\
\hline
\multicolumn{2} {|c||}{\multirow{ 2}{*}{\textbf{TOTAL}}}    & {\textbf{3070}}    &  {\textbf{1990}}   &  {\textbf{593}} &   &  {\textbf{56}}  &  \textbf{< 0.1\% [178]} & {\textbf{1}}     &   \\
  &  & {\textbf{100\%}}   & {\textbf{65\%}}  & {\textbf{19\%}} &  & {\textbf{2\%}}  &  & {\textbf{< 0.1\%}}  &    \\
\hline
\end{tabular}
\end{table*}
}
{\scriptsize
\begin{table*}[h]
    \centering
    \caption{Data of Table \ref{tab:impactLLVMalive}, aggregated by tool and by compiler bug.}
        \SPACEHACK{-3mm}
    \label{tab:impactLLVMotherbugs}
\begin{tabular}{|c||c||ccc|c|c|}
\hline
{\multirow{ 2}{*}{\textbf{TOOL}}} &  \multirow{ 2}{*}{\textbf{BUGS}}  & \multicolumn{3} {c|} {\textbf{1) BUGGY LLVM CODE}} & {\textbf{2) BINARY DIFFS}} &   {\textbf{3) RUNTIME DIFFS}} \\
    & & \textit{reached} & \textit{triggered} & \textit{(precise)}  & \textit{packages} &  \textit{test diffs} \\
\hline
 Alive & 8 & 7 & 2 & (0) & 2 & 0   \\
 User-reported & 10 & 8 & 4 & (0) & 2 & 1  \\
\hline
\end{tabular}
\end{table*}
}

The impact data for the \numAliveBugs bugs discovered by Alive and our sample of
\numUserBugs user-reported bugs is reported in Tables~\ref{tab:impactLLVMalive}
and~\ref{tab:impactLLVMotherbugs}.

\myparagraph{Alive Bugs.} The average impact of the Alive bugs appears
much more limited than the one of the Csmith and EMI bugs. The buggy
compiler code is reached twice less often for the Alive bugs,
and \numNeverTriggeredAliveBugs/\numAliveBugs Alive bugs never
trigger, compared to only \numNeverTriggeredCsmithEMIBugs/20 for
Csmith and EMI. The bug impact profile for Alive is actually closer to
the one for Orange: Alive bugs trigger in the context of corner cases
(typically overflows) of possibly complex arithmetic, which do not
appear so often in real-world code. However, two Alive bugs led to
binary differences affecting 172 packages and a higher total
proportion of functions than the bugs reported by any fuzzer.

\myparagraph{User-reported Bugs.} Contrary to fuzzer-found bugs,
user-reported ones were discovered by spotting miscompilations in the
code of a real application. Our results tend to show that this does
not make user-reported bugs likelier to trigger when compiling other
applications. On the contrary, different binaries are only spotted for
\numBinaryDiffsUserBugs/\numUserBugs bugs and \percentBinDiffByBuildUser of the package builds,
far below the values of 11/20
and \percentBinDiffByBuildCsmithEMI reached by Csmith and EMI. Similar to fuzzer-found bugs, we have observed a single test failure for package \textit{leveldb} with bug 27903 (the technical details of this miscompilation are also available in the appendix \S\ref{sec:miscomp-details}), but however no test failure when compiling \sqlite. Manual inspection also suggests a low bug impact as both bugs \#27903 and \#33706 require a very specific interaction between some compiler optimisations to happen for possibly impacting the generated code.

\subsection{Correlation between Bug Severity and Impact}\label{sec:severity-and-impact}
{\scriptsize
\begin{table*}[t]
    \centering
    \caption{Results for our 45 LLVM bugs over our \numPackages Debian packages, aggregated by bug severity.} 
        \SPACEHACK{-3mm}
    \label{tab:impactLLVMtotal}
\begin{tabular}{|cc||c||cc|cc|c|}
\hline
\multicolumn{2} {|c||}{\textbf{BUG AGGREGATE}} & \textbf{PACKAGES} & \multicolumn{2} {c|} {\textbf{1) BUGGY LLVM CODE}} & \multicolumn{2} {c|} {\textbf{2) BINARY DIFFS}} &  {\textbf{3) RUNTIME DIFFS}} \\
\textit{severity} &  \textit{{\#bugs}} & \textit{successful builds}  & \textit{reached} & \textit{triggered} & \textit{packages} & \textit{functions} & \textit{test diffs}   \\
\hline
\hline
\multirow{ 2}{*}{enhancement} &  \multirow{ 2}{*}{{8}}  &  2449 & 2054 & 968 & 84 & < 0.1\% [279]& 0  \\
& & {{100\%}}   &  84\%  & 40\%  & 3\% & & 0\%   \\
\hline
\multirow{ 2}{*}{normal} &  \multirow{ 2}{*}{{33}} & 10151 & 5608 & 1941 & 611 & 0.3\% [8130] & 2   \\
& & {{100\%}}   &   55\%  & 19\%  & 6\% & & < 0.1\%      \\
\hline
\multirow{ 2}{*}{release blocker} &   \multirow{ 2}{*}{{4}} & 1226 & 1191 & 294 & 10 & < 0.1\% [16] & 0  \\
&  & {{100\%}}   & 97\% & 24\%  & 0.8\% & & 0\%     \\
\hline
\end{tabular}
\end{table*}
}

Table~\ref{tab:impactLLVMtotal} aggregates the impact data by the
severity level that bugs were assigned to on the Clang/LLVM bug
tracker. We expected a higher impact to be associated with bugs with
higher severity, but no clear trend emerged in practice and some
numbers are even counter-intuitive: the buggy compiler code is reached
almost twice as often at the \textit{enhancement} level than at the
more severe \textit{normal} level, while different binaries are
produced three to six times less often at the \textit{release blocker} level
than at the two lower levels. While the confidence in these results
would be increased by bigger samples at the \textit{enhancement} and
\textit{release blocker} levels, the bug severity appears to be a bad
predictor of the practical bug impact.

\subsection{Discussion}\label{sec:discussion}

We now discuss the main results of our study. As with any experimental
study, these results are subject to several threats to both internal
and external validity, which we detail in \S\ref{sec:threats}.

Our top-level findings include
that the code associated with each fuzzer-found bug is reached always at least once and typically quite frequently
when compiling our set of real-world applications, that our bug
conditions trigger, and almost half of the bugs result in binary-level
differences for some packages.  However, these differences affect a
tiny fraction of the functions defined in the application's code and
they cause a couple of test failures in total, one in \sqlite (due to
the miscompilation of its test bed) and one in the \zsh
shell. Regarding the user-reported and Alive bugs, there exist bugs for which the associated code is {never} reached during the compilation of our set of packages. In addition, these bugs have their conditions
triggered less frequently on average, and lead to a single test failure in the
\leveldb database system. The manual analysis of a selection of the binary-level
differences caused by some of our compiler bugs,
both fuzzer-found and non fuzzer-found, shows that either these binary differences have no impact over the semantics of the application, or that they would require very specific runtime circumstances in order to corrupt the internal state of the application and propagate to its visible outputs.

To sum up, considering the two objectives of this study defined in the introduction (\S\ref{sec:intro}), our major take-aways are that (a) miscompilation bugs (whether fuzzer-found
or not) in a mature compiler infrequently impact the correctness and reliability of
real-world applications, as encoded by their test suites and (in a few cases) based on our manual
analysis, and (b) fuzzer-found miscompilation bugs appear to have at
least as much impact as bugs found via other sources, particularly
bugs reported by the
compiler users because they were affecting real code.

In safety-critical software, where avoiding the triggering of any
single bug can make the difference between life and death, using a compiler that has been intensively tested with a
dedicated fuzzer appears to be a fundamental requirement. Of course,
one could argue that in such a case, using a formally verified
compiler would be even better. However, such compilers still do not
offer the same language features and performance as traditional
compilers, and they are typically not verified end-to-end, enabling
fuzzers to find bugs in their non-verified parts \cite{csmith}.  In
non safety-critical software, where one can tolerate a certain level
of unreliability in the application, in exchange for a reduced
software development cost, our study supports the claim that
miscompilation bugs in the compiler seem unlikely to cause a dramatic
decrease of the application's reliability level, at least when using a
sufficiently mature compiler platform.

It might appear counter-intuitive that our study does not reveal any clear difference with respect to the practical impact of the bugs, depending on whether they were discovered by a fuzzer compiling artificial code or a human user compiling real code. Indeed, our initial intuition, highlighted in the introduction (\S\ref{sec:intro}), was on the contrary that bugs found while compiling real code might trigger more often over applications in the wild than bugs discovered while compiling artificial code. Given the rather limited impact of all the bugs considered in this study, one hypothesis to explain this unexpected result could be that, in a mature compiler like Clang/LLVM, all the bugs affecting code patterns that are frequent in real code have already been fixed, so that the remaining bugs are corner cases that do not appear more frequently in real code than in artificial code.

Finally, as demonstrated by the huge list of unfixed compiler bugs in the
Clang/LLVM bug repository,\footnote{More than 10,000 bug reports remain
  open as of August 2019.} developers simply do not have time to
resolve all the issues. While these issues seem to be prioritised
based solely on information provided by the bug reporters and the intuition of the developers, our study suggests
that this might not to be a good indicator of the actual impact that
the bugs have over the correctness and reliability of a wide and various sample of applications. This supports the idea that alarmist bug reports or bad intuitions from the developers can end up delaying the fixing of high impact bugs, calling for further research on how to better prioritise compiler bug reports.

%% file: 6threats.tex

\myparagraph{Threats to Internal Validity.}
A first class of threats to the validity of our experimental results
arise because the software artifacts that we used, including the shell
scripts, warning-laden fixing patch, compilers, package development
framework and system tools, could be defective. However, we have
crosschecked our results in several ways. The data computed by the
shell scripts were verified by hand for a small number of packages and
bugs. A sample set of the warning-laden fixing patches were randomly
selected and reviewed by a member of the team who had not been involved
in producing them. Each warning-laden compiler was tested over the
miscompilation samples provided on the bug tracker. Any possible
contradiction between the results of Stages 1 and 2, where no
potential fault would trigger but the produced binaries would be
different, was investigated. A sample of the occurring suspicious
behaviours like build failures or triggered bugs leading to similar
binaries were investigated for a satisfiable explanation. The test
logs were hand-checked and the test runs were repeated several times
in clean environments for a dozen bug and package
pairs, picked at random, and also in all the situations where the binaries produced
divergent test results or where the test beds crashed. All these
sanity checks succeeded.

Another threat is that some of the fixing patches from the Clang/LLVM
repository could be incorrect. However, this seems improbable given
that all the associated bug reports have been closed, without being reopened,
for at least several months and typically for years.

Our results might also have been affected by the unreproducible build
process of some Debian packages. However, this is very unlikely as all
the used packages were selected among the list of reproducible packages
provided by Debian. Moreover, each of the selected packages was built
twice and checked for any unreproducible behaviour. Analysing unreproducible packages should also have led to contradictions between
Stage 1 and Stage 2, i.e. binary differences would have been spotted without any bug warning being issued first during compilation, but none were
detected.

Aggressively leaving out all the assembly operands during binary
comparison may have led to undercounting the number of different
assembly functions at Stage 2. However, the insight that we gained along all our experiments about the studied bugs suggests that none of them would only affect the operands
in the binaries produced by the buggy compiler.

The effectiveness of a part of our dynamic binary analysis is impacted
by the thoroughness of the test suites, and in particular whether
these test suites are deemed thorough enough to act as a proxy for
typical real-world usage. A study of test coverage for a sample of our
packages revealed that their test suites achieve a median 47\%
statement coverage, peaking at 95\%, which appears high enough for not
attributing the absence of test suite failures to very poor test
coverage.  Moreover, we repeated our dynamic binary analysis with the
\sqlite application, relying on its highly thorough free test suite
(more than 98\% coverage of \sqlite's 151K statements), which detected
a single failure, related to the miscompilation of part of its test
suite rather than the application code itself. This being said, test
suites are clearly a weak proxy for all possible
behaviours of an application.

Finally, if the test suites of some of our packages were used to
regularly exercise Clang/LLVM, it might potentially have skewed results in
two ways.  First, if compiling a package and running its test suite
were part of the regression testing process for Clang/LLVM, the test suite might
have detected miscompilation bugs that were fixed without any trace.
In other words, some miscompilation bugs might have affected
application correctness, but fuzzers would not have had a chance to
detect them, as the bugs were fixed during the compiler development process.  To
avoid this, we have selected packages that are not part of the
extended Clang/LLVM test suite.\footnote{\url{https://llvm.org/docs/TestSuiteGuide.html}}
Second, if the developers of a package regularly compile it using Clang/LLVM,\footnote{While GCC is default in Debian, Clang/LLVM might notably be used for packages deployed on other platforms.} they might work around miscompilation bugs by changing the
package code and/or its test suite, and thus our approach would
miss that the compiler bug affected that package at some point in
time.\footnote{As an example, see this Mozilla Firefox work-around (\url{https://bugzilla.mozilla.org/show_bug.cgi?id=1481373}) for an LLVM miscompilation bug (\url{https://bugs.llvm.org/show_bug.cgi?id=38466}).}
However, it is unlikely that more than a fraction of
packages are subject to such an approach to avoid miscompilation bugs.
Furthermore, some of our bugs did not make it into released versions
of the compiler, and it is unlikely that too many developers use
non-released versions to compile their packages.  Also, the
development of some of our packages is still active now, in 2019. All
the package code that was developed a reasonable time after each of
our bugs was fixed has a small chance to have ever been compiled with
a build of Clang/LLVM affected by the bug.  Conversely, some of our
packages are old and their development had already been discontinued
years before our compiler bugs were introduced or reported.
For example, the last version of \libgthreed was released in 2009, while
the bulk of our compiler bugs have lifespans corresponding to short
periods between 2012 and 2016.  Finally, regression test suites usually get
stronger as time goes by, so that the test suites of the
current versions of the packages, used in our experiments, might still
find miscompilations that were not found if some older test suites were run
over packages compiled using Clang/LLVM in the past.

\myparagraph{Threats to External Validity.} Common to all empirical
studies, this one may be of limited generalisability. To reduce this
threat, we performed our experiments over \numPackages diverse
applications from the well-known Debian repository, including some
very popular ones like \apache and \grep, totalling more
than \numLOCMillion millions lines of code. We have also investigated
45 historical Clang/LLVM bugs. They include a sample of about 12\% of
all the fixed miscompilation bugs reported by the studied fuzzers in
Clang/LLVM. Moreover, our sampling strategy excluded many bugs that
are likely to have lower practical impact due to their reliance on
specific and uncommon compiler flags.

Our study was performed on Clang/LLVM, which is a mature and widely
used compiler platform. Experimenting with other and notably less
mature compilers, e.g. for emerging languages like Rust, might produce
different results, as higher impact bugs may be more likely in such
compilers.

%% file: future.tex

We discuss now what we believe to be the three main future research directions to evaluate even better the impact of compiler bugs over applications in the wild.

First, this study introduces both a conceptual methodology and an
experimental framework (available as a part of the paper's associated
artifact), which pave the way for others to perform similar
evaluations of the impact of miscompilation bugs, possibly replicating
our results, studying additional bugs or targeting other compilers,
notably compilers less mature than Clang/LLVM.

Second, as a simplifying hypothesis, we have limited our analysis of
compiler bugs to their impact over the correctness and reliability of
the compiled application. In practice, these bugs could also impact
non-functional properties, such as efficiency or security. Developing
effective operational metrics to measure this additional impact is a
challenging task that could improve even more our understanding of the
risks induced by compiler bugs. In particular, it would be interesting
to estimate the attack surface area provided by miscompilation bugs,
as it has been shown that any such bug could potentially be used by
attackers to introduce backdoors in an
application~\cite{regehr:backdoors15,compiler-backdoor:blog15}.

Third, Stage~3 of our study looks for runtime divergences caused by
miscompilation bugs, using either the application regression test
suite or inputs that we have crafted manually. Proving that no
behavioral differences could ever happen between the binaries produced
by the buggy and fixed compilers is nothing else than establishing
that two pieces of code are semantically equivalent, which is
undecidable. However, with a significant engineering and computation
effort, one could probe these binaries more thoroughly, by using an
automated test generation approach (such as symbolic
execution \cite{symex:cacm}) to produce inputs reaching the
miscompiled sections in the application code, and by leveraging
efficient oracles (such as multi-version execution~\cite{varan}) to
detect runtime failures induced by the miscompilation.

%% file: 7related.tex

\myparagraph{Understanding Compiler Bugs.}
A recent empirical study provides an in-depth analysis of bugs in the
GCC and LLVM compilers~\cite{compiler-bugs:issta16}, focusing on
aspects such as the distribution of bugs across compiler components,
the sizes of triggering test cases associated with bug reports, the
lifetime of bug reports from filing to closing, and the
developer-assigned priority levels for bugs and how these correlate to
compiler components.  The study is complementary to ours: beyond a
discussion of bug priorities, it is not concerned with the extent to
which compiler bugs affect real-world applications, and it does not
focus on whether the bugs under analysis are miscompilations, nor
whether the bugs were found in the wild or via automated tools such as
fuzzers.

Another empirical study~\cite{Chen2016} compares
the equivalence modulo inputs and differential testing approaches to
compiler testing (see below).  A ``correcting
commits'' metric is proposed that helps to identify distinct compiler
bugs from failing tests.  Otherwise the focus of
the study is on understanding the testing techniques themselves,
rather than understanding the real-world impact of the bugs they find.

The paper associated with the Csmith tool~\cite{csmith} focuses to
some degree on understanding compiler bugs, \eg identifying the most
buggy files (according to Csmith-found bugs) in versions of GCC and
LLVM at the time.  This analysis \emph{does} distinguish between wrong
code bugs and crash bugs, but is simply concerned with whether bugs
exist, rather than (as in our study) whether they affect real-world
applications.  Two projects associated with Csmith, on automatically
reducing test cases that trigger fuzzer-found bugs~\cite{creduce}, and
on ranking reduced test cases in a manner that aims to prioritise
distinct bugs~\cite{chen2013}, are important for
understanding the root causes of fuzzer-found bugs, but do not
directly shed light on how likely such bugs are to be triggered by
real applications.

Bauer et al.~\cite{regehr:backdoors15} discuss exploiting compiler bugs to create software backdoors, and
show 
a proof-of-concept backdoor based on the Clang/LLVM miscompilation bug \#15940 found by the Orange3 tool.  The possibility of code written
to maliciously exploit a known miscompilation bug
presents a compelling argument that miscompilations matter even though
they may not otherwise affect real-world code.
%
An informal online collection of anecdotes about compiler bugs found in the wild also 
makes for interesting
reading.\footnote{\url{http://wiki.c2.com/?CompilerBug}}    

\myparagraph{Automated Compiler Testing.}
The idea of randomly generating or mutating programs to induce errors
in production compilers and interpreters has a long history, with
grammar- or mutation-based fuzzers having been designed to test
implementations of languages such as COBOL~\cite{cobol-testing62},
PL/I~\cite{Hanford70}, FORTRAN~\cite{compiler-bugs:fortran}, Ada and
Pascal~\cite{Wichmann98}, and more recently
C~\cite{csmith,emi,athena,hermes,orange3,orange4,yarpgen},
JavaScript and PHP~\cite{fuzzing-fragments}, Java
byte-code~\cite{Chen2016}, OpenCL~\cite{clsmith},
GLSL~\cite{OpenGLtesting:met16,compiler-bugs:PACMPL17}
and C++~\cite{compiler-bugs:issta16} (see also two surveys on the
topic~\cite{Boujarwah97,Kossatchev2005}).  Related approaches have
been used to test other programming language processors, such as
static analysers~\cite{csmithFramaC}, refactoring
engines~\cite{testing-refactorings}, and symbolic
executors~\cite{symex-test}.  Many of these approaches are either
geared towards inducing crashes, for which the test oracle problem is
easy.  Those that can find miscompilation bugs do so typically via
\emph{differential testing}~\cite{mckeeman:diff-test}, whereby
multiple equivalent compilers, interpreters or analysers are compared
on random programs, or via \emph{metamorphic
testing}~\cite{metamorphic98,segura2016survey}, whereby a single tool
is compared across equivalent programs, or generating programs with
known expected results.

Regarding the fuzzers of our study, Orange3 takes the approach of
generating programs with known results~\cite{orange3};
Csmith~\cite{csmith} and Yarpgen~\cite{yarpgen} are intended to be
applied for differential testing; while the \emph{equivalence modulo
inputs} family of tools~\cite{emi,athena,hermes} as well as
Orange4~\cite{orange4} represent a successful application of
metamorphic testing (earlier explored with only limited
success~\cite{compiler-bugs:metamorphic-testing}).

A recent non-fuzzing compiler testing technique
involves \emph{skeletal program enumeration}: exhaustively enumerating
all programs (up to $\alpha$-renaming) that have a particular
control-flow skeleton~\cite{skeletal-program-enum}.  This technique is
geared towards finding compiler crashes rather than miscompilations,
so the bugs that it finds are not relevant for a study such as
ours.  

\myparagraph{Compiler Verification.}
A full discussion of compiler verification is out of scope for this
paper, but we mention CompCert~\cite{CompCert} as the most notable
example of a \textit{formally verified} compiler. 
CompCert provides an incomparable level of reliability: intensive
fuzzing via Csmith and EMI techniques have not discovered any bugs in
the verified parts of the code base~\cite{csmith,emi}
(as should be expected for a formally verified piece of software). 
One might think 
that a verified compiler should make the question of whether compiler
bugs matter irrelevant by eliminating bugs completely.  However,
CompCert still faces some major limitations, such as incomplete
language support (including no support for C++) and a less mature set
of optimisations compared with \eg GCC or LLVM.  A compromise is to
verify certain parts of a compiler, an approach taken by
Alive~\cite{alive}, whose bugs we have included in this study.

%% file: 8conclusion.tex
Compiler fuzzing tools have proven capable of finding hundreds of
errors in widely-used compilers such as GCC and LLVM, but very little
attention has been paid to the extent to which they affect real-world
applications. In this first empirical study investigating these
questions, we have shown that almost half of the fuzzer-found bugs in
our sample propagate to the compiled binaries of real-world
applications, but affect a very small number of functions and
lead to a small number of test suite failures. At the same time, our study suggests that fuzzer-found
compiler bugs appear to have at least as much impact as bugs found via
other sources, particularly bugs reported by the
compiler users because they were affecting real code. The manual analysis of a selection of the syntactic changes
 caused by some of our bugs (fuzzer-found and non fuzzer-found) in package assembly code, shows that either these changes have no semantic impact or that they would require very specific runtime circumstances to trigger an execution divergence.

%% file: appendix.tex

\subsection{Three Miscompilation-induced Test Failures}  \label{sec:miscomp-details}

\myparagraph{Test Failure in Z Shell (caused by EMI bug \#29031).} Z shell (zsh) is a popular Unix command-line interpreter, whose default test suite covers 54\% of the its 54K statements. The buggy compiler miscompiles the following section of code in the function `paramsubst' of the `subst.c' file, by wrongly hoisting the statement \lstinline{*t = sav;} on line 16 outside of the while loop:
\begin{lstlisting}[basicstyle=\ttfamily\footnotesize,language=C,numbers=left, stepnumber=1, numberfirstline=false, firstnumber=1, xleftmargin=4ex, numberstyle=\tiny]
sav = *t; // s and t respectively point to the 1st and 2nd `:' of "${(g:o:)foo}"
*t = 0;
while (*++s) {
    switch (*s) {
        case 'e':
            getkeys |= GETKEY_EMACS;
            break;
        case 'o':
            getkeys |= GETKEY_OCTAL_ESC;
            break;
        case 'c':
            getkeys |= GETKEY_CTRL;
            break;
        default:

            *t = sav; // wrongly hoisted outside the while loop on line #3
            goto flagerr;
    }
}
\end{lstlisting}
This miscompilation makes the test case named `{\$\{(g)...\}}' in the test script file `D04Parameter04.ztst' fail. When this test is executed with correctly compiled code, the code section above is processed, starting with variables \lstinline[basicstyle=\ttfamily\footnotesize]{s} and \lstinline[basicstyle=\ttfamily\footnotesize]{t} pointing to the first and the second `:' in a string object "\lstinline[basicstyle=\ttfamily\footnotesize]|${(g:o:)foo}|".  On line~1, the character `:' pointed by \lstinline[basicstyle=\ttfamily\footnotesize]{t} is saved for future restoration. Line~2 replaces the second `:' of "\lstinline[basicstyle=\ttfamily\footnotesize]|${(g:o:)foo}|" by a \lstinline[basicstyle=\ttfamily\footnotesize]{NUL} character in the string object.
The loop condition then makes \lstinline[basicstyle=\ttfamily\footnotesize]{s} move one character forward in the string object. As the new character pointed by \lstinline[basicstyle=\ttfamily\footnotesize]{s} is an `o', it is not NUL, so that the loop is entered and the `o' case of the switch statement is executed. The loop condition then makes \lstinline[basicstyle=\ttfamily\footnotesize]{s} move again one character forward in the string object. The new character pointed by \lstinline[basicstyle=\ttfamily\footnotesize]{s} is now NUL and the loop is exited. However, this last operation is not executed in this way when the code is miscompiled. Indeed, by hoisting the statement which restores the character pointed by \lstinline[basicstyle=\ttfamily\footnotesize]{t} in the string object to a `:' (line~16) outside the while loop, the miscompilation will cause the new character pointed by \lstinline[basicstyle=\ttfamily\footnotesize]{s} not to be NUL anymore, but `:', so that the loop is not exited and the default case of the switch statement is executed, leading to an error being reported.

\myparagraph{Test Failure in SQLite (caused by CSmith bug \#13326).} The buggy compiler
miscompiles the following line of code in the test bed of test \textit{incrblob-2.0.4}, by generating erroneous binary
code to compute an 8-bit unsigned integer division, part of the
remainder evaluation:
\begin{lstlisting}[basicstyle=\ttfamily\footnotesize,language=C]
zBuf[i] = zSrc[zBuf[i]%(sizeof(zSrc)-1)]; // RUNTIME: sizeof(zSrc)=79, zBuf[i]=232
\end{lstlisting}
When test \textit{incrblob-2.0.4} is then run, the previous code mistakenly returns 254 when computing the remainder of 232 by 78,
causing \lstinline[basicstyle=\ttfamily\footnotesize]{zBuf[i]} to take a garbage value \lstinline[basicstyle=\ttfamily\footnotesize]{zSrc[254]}.
As the test  uses \lstinline[basicstyle=\ttfamily\footnotesize]{zBuf}
 to exercise a database handled by \sqlite, it ends up reporting an error when it detects a value
 different from the one it expected.

\myparagraph{Test Failure in LevelDB (caused by user-reported bug \#27903).} LevelDB is an open source database management system currently being developed by Google.
The buggy compiler may miscompile the function `MakeFileName' from the `filename.cc' file by generating wrong code for the string concatenation operation \lstinline{name + buf} on line~7:
\begin{lstlisting}[basicstyle=\ttfamily\footnotesize,language=C,numbers=left, stepnumber=1, numberfirstline=false, firstnumber=1, xleftmargin=4ex, numberstyle=\tiny]
static std::string MakeFileName(const std::string& name, uint64_t number,
const char* suffix) {
    char buf[100];
    snprintf(buf, sizeof(buf), "/%06llu.%s",
             static_cast<unsigned long long>(number),
             suffix);
    return name + buf; // problem when compiling the '+' operator function
}
\end{lstlisting}
This miscompilation may make one or several default tests fail. The bug affects the  enhancement of stack coloring data flow analysis, which is designed to reduce the runtime memory consumption of the compiled executable code. The function above takes the path to a directory (\lstinline[basicstyle=\ttfamily\footnotesize]{name}) as a parameter and is supposed to return the full path of a file under that directory. On line~7, the string \lstinline[basicstyle=\ttfamily\footnotesize]{name} and the string \lstinline[basicstyle=\ttfamily\footnotesize]{buf} are concatenated to form the full path of the file. However, the buggy compiler miscalculates the lifetime of the \lstinline[basicstyle=\ttfamily\footnotesize]{buf} variable, so that the compiled binary will mistakenly use the memory zone pointed by \lstinline[basicstyle=\ttfamily\footnotesize]{buf} to store its memory address, overwriting thus the content of the variable by its address. If the first byte of that address is all zero, the function `MakeFileName' will then return a string whose value is the same as \lstinline[basicstyle=\ttfamily\footnotesize]{name}, the directory path. When LevelDB then tries to write to the file returned by `MakeFileName', it ends up reporting an error when it detects that the expected file is actually a directory.



\subsection{Salient Observations made by Manually Inspecting Syntactic Binary Differences} \label{sec:manual-anal}

\myparagraph{CSmith Bug \#11964.}
\begin{itemize}[leftmargin=*]
\item \textit{Bug description}: During the transformation of LLVM IR into x86 assembly code (represented within the compiler's data structures by directed acyclic graphs (DAG)), the bug causes an IR node representing a decrementation by one for a long integer value to be miscompiled, in case the node has more than one other node using it in the DAG. This miscompilation makes the operation to be possibly performed twice in the x86 code.
\item \textit{Manual inspection}: We have reviewed all the syntactically affected functions in the two impacted packages, namely \textit{s3ql} and \textit{simplejson}. All the syntactic differences in the functions seem to be related to a memory management macro \lstinline[basicstyle=\ttfamily\footnotesize,language=C]{Py_DECREF(PyObject * o)}, which is used to keep track of the reference count of a Python object and deallocate the object once the reference count decreases to zero. While we did see internal states corrupted, these corruptions do not impact the global semantics of the packages. For example, the values stored in the x86 RAX register may differ by one at the end of the function \lstinline[basicstyle=\ttfamily\footnotesize,language=C]{join_list_unicode} of package \textit{simplejson}, but these values are not used again until the function returns.
\item \textit{Generalised impact estimation}: While posing a clear threat of a miscompilation failure, this bug requires specific conditions to be met to trigger, so that it affects a highly limited amount of functions in our packages. For this reason, we rate its impact as \textit{low}.
\end{itemize}

\myparagraph{CSmith Bug \#27392.}
\begin{itemize}[leftmargin=*]
\item \textit{Bug description}: The bug affects the loop
unrolling optimisation in the case where the loop is unrolled by a
given factor at compile-time, but it is not possible to evaluate
before runtime whether the total number of times the loop
must be executed is a multiple of the loop unrolling factor. To deal
with this case, the compiler produces extra target code to find out at
runtime how many iterations were possibly left over and execute
them. This extra target code evaluates at runtime the number of times
the non-unrolled loop must be executed and stores it into an unsigned
integer counter variable. However, the case where this number
overflows the counter variable is not properly handled, making the
loop to be executed a wrong number of times when this case occurs.
\item \textit{Manual inspection}: We inspected manually a dozen of the many syntactic binary differences induced by this bug, without being able to trigger any runtime divergence. The main obstacle for such a divergence to occur is that the binary difference should affect the loop unrolling management code of a loop able to iterate more times than the largest representable unsigned integer value, which was not the case for any of the affected loops we investigated. For example, the while loop in the main function of the \textit{boxes} package (\textit{boxes.c}, line 1661) can be run at most twice, while the for loop in the hash table API of the \textit{conntrack-tools} package (\textit{hash.c}, line 45) can be run at most a number of times equal to the largest representable signed integer value: \begin{lstlisting}[basicstyle=\ttfamily\footnotesize,language=C]
struct hashtable *hashtable_create(int hashsize, ...
for (i=0; i<hashsize; i++) INIT_LIST_HEAD(&h->members[i]);
\end{lstlisting}
\item \textit{Generalised impact estimation}: This bug causes syntactic binary differences in many packages, because its fixing patch systematically affects the syntax of the loop unrolling management code produced by the compiler, and because such management code will often be generated by the compiler in the expected presence of loops in the package. However, such syntactic differences are typically not the witnesses of miscompilation failures, because it is very unlikely that the application contains loops whose intended behaviour is to iterate more times than largest representable unsigned integer value. As a consequence, we rate the bug impact as \textit{very low}.
\end{itemize}

\myparagraph{EMI Bug \#26323.}
\begin{itemize}[leftmargin=*]
\item \textit{Bug description}: During one optimisation pass, the compiler may merge a disjunction between two simple integer comparisons into a single comparison. For example, \lstinline[basicstyle=\ttfamily\footnotesize,language=C]{(n==2 || n==3)} would be merged into \lstinline[basicstyle=\ttfamily\footnotesize,language=C]{(n & ~1)==2}. During a later optimisation pass,  the compiler may analyse the IR to detect if such a merger has occurred before, in order to undo it and apply another code transformation instead. However, the analysis to locate merger occurrences in the IR is flawed, leading the compiler to trigger the merger undoing process over code where no merger was actually performed, possibly corrupting the code's semantics.
\item \textit{Manual inspection}: Despite the exhaustive manual inspection of several syntactic binary differences induced by the bug, we were not able to trigger any runtime divergence. The syntactic differences are caused by the fact that the buggy compiler applies the merger undoing process to some code, while the fixed compiler does not. However, manual inspection revealed that applying the merger undoing process to the code that we inspected was actually legitimate and would preserve the semantics of the package. For example, the differences spotted in the \textit{libraw} (\textit{dcraw\_common.cpp}, line 6308) and \textit{art-nextgen-simulation-tools} (\textit{art\_SOLiD.cpp}, line 144) packages are caused by the legitimate undoing by the buggy compiler of the two following mergers: \begin{lstlisting}[basicstyle=\ttfamily\footnotesize,language=C,escapeinside={(*}{*)}]
!(id==308 || id==309) (*$\rightarrow$*) !((id & ~1)==308)
!(k==6 || k==7) (*$\rightarrow$*) !((k & ~1)==6)
\end{lstlisting}
\item \textit{Generalised impact estimation}: While the bug is caused by a flawed merger locator, the fixing patch replaces this locator by a correct but also overly strict one. As a consequence, the fixed compiler will not apply merger undoing mistakenly any more, but it will also miss many legitimate opportunities to do so which were correctly seized by the buggy compiler. A detailed analysis shows that the fixed compiler applies the following code transformation (where $c_1$ and $c_2$ are positive integer constants):
\begin{lstlisting}[basicstyle=\ttfamily\footnotesize,language=C,escapeinside={(*}{*)}]
(x & ~2(*$^{c_1}$*))==2(*$^{c_2}$*) (*$\rightarrow$*) (x==2(*$^{c_2}$*) || x==(2(*$^{c_2}$*)|2(*$^{c_1}$*))) (*where $c_1 \neq c_2$ *)
\end{lstlisting}
while the buggy compiler has notably the ability to apply the following (more general but still correct) transformation:
\begin{lstlisting}[basicstyle=\ttfamily\footnotesize,language=C,escapeinside={(*}{*)}]
(x & ~2(*$^{c_1}$*))==(*$c_2$*) (*$\rightarrow$*) (x==(*$c_2$*) || x==((*$c_2$*)|2(*$^{c_1}$*))) (*where the bit position ${c_1}$ of $c_2$ is 0*)
\end{lstlisting} This situation prevents many syntactic binary differences from affecting the application's semantics. Considering also the tiny frequency of the syntactic binary differences induced by this bug, we rate the bug impact as \textit{very low}.
\end{itemize}

\myparagraph{EMI Bug \#29031.}
\begin{itemize}[leftmargin=*]
\item \textit{Bug description}: During one optimisation pass, some operations that use or assign a variable might be hoisted to an earlier point in the code of the current function (referred to as the hoisting point), provided that the candidate operation for hoisting is executed on all execution paths from the hoisting point to a function exit point. The bug is that checking for this last condition to hold may wrongly allow hoisting when the hoisting point is within a loop which contains a path where the candidate operation is not executed before exiting the loop. When this arises, the compiler will produce code that mistakenly execute the candidate operation when this path is executed.
\item \textit{Manual inspection}: We were unable to trigger a runtime divergence for this for this bug, despite notably testing different loop execution scenarios in the \textit{aragorn} package (\textit{aragorn1.2.38.c}, line 11503). A deeper analysis of the fixing path reveals that it deactivates hoisting as soon as the hoisting point is within a loop and does not live in the same basic block as the candidate operation, even if the candidate operation is always executed within the loop and even if executing this operation during a path where it should have not been executed will not affect the code semantics.
\item \textit{Generalised impact estimation}: Given the careful approach taken in the fixing patch, a number of the binary syntactic differences that it triggers will not impact the application's semantics and thus not be the witnesses of a miscompilation failure. For the binary syntactic differences which cause actual failures, the failure will only impact executions that follow one or some specific paths within the loop. For these reasons, we rate the impact of the bug as \textit{low}.
\end{itemize}

\myparagraph{EMI Bug \#30935.}
\begin{itemize}[leftmargin=*]
\item \textit{Bug description}: During one optimisation pass, the compiler can hoist instructions out of a loop if these instructions are invariant from one loop iteration to another. In case the hoisted instruction is a division, it may occur that the loop entry condition, which was guarding the division execution before hoisting, was preventing the division to be performed with a zero divisor. After hoisting the division out the loop, this division is performed even when the loop entry condition is false, possibly making the program mistakenly crash with a zero divisor error.
\item \textit{Manual inspection}: We reviewed each of the three syntactic binary differences induced by this bug. The differences in the \textit{nauty} (\textit{genrang.c}) and \textit{pixbuf} (\textit{pixops.c}, line 316) packages revealed to be particularly hard to analyse: either there were no clear tracks of a division hoisting in the assembly code, or such tracks were present, but it was totally unclear which statement in the C/C++ source could have been compiled into assembly code involving this division. The hoisting was much more obvious to understand in the \textit{infernal} package (\textit{esl\_histogram.c}, line 957), but it was also clear that the affected code (detailed below) prevented any division by zero:
\begin{lstlisting}[basicstyle=\ttfamily\footnotesize,language=C,escapeinside={(*}{*)}]
if (maxbar > 0) units = ((maxbar-1)/ 58) + 1;
else            units = 1;
// The previous definition of "units" prevents it to be 0 at hoisting point
...
// Hoisting point
for (i = h->imin; i <= h->imax; i++)
    { ...
      num = 1+(lowcount-1) / units; // Hoisted division
\end{lstlisting}
\item \textit{Generalised impact estimation}: The fixing patch for this bug takes a conservative approach and disables hoisting of expressions containing divisions by anything other than non-zero constants. Because of this, some of the syntactic binary differences may be false alarms similar to the one above. However, it appears not unlikely that, in other situations, the affected code would allow the divisor to be zero at loop entrance, while the loop condition would be false, leading to a real miscompilation failure. Nevertheless, considering the tiny frequency of the binary differences induced by this bug, we rate its impact as \textit{low}.
\end{itemize}

%% file: 9acknowledgment.tex
\section*{Acknowldegments}


Our experimental infrastructure reuses parts of a framework
developed by Syl\-ve\-stre Ledru (Mozilla France) and Lucas Nussbaum
(Universit\'e de Lorraine), whom we thank for their support. We also
thank John Regehr (University of Utah) for his input regarding the bugs found by the
Csmith tool and Pritam Gharat (Imperial College London) for proofreading the paper.  Finally, we thank the anonymous reviewers of this paper for their valuable comments.

This work was supported by EPSRC projects EP/R011605/1 and EP/R006865/1.


%% file: compiler-bugs.bbl

\newcommand{\noopsort}[1]{} \newcommand{\printfirst}[2]{#1}
  \newcommand{\singleletter}[1]{#1} \newcommand{\switchargs}[2]{#2#1}
\begin{thebibliography}{41}


\ifx \showCODEN    \undefined \def \showCODEN     #1{\unskip}     \fi
\ifx \showDOI      \undefined \def \showDOI       #1{#1}\fi
\ifx \showISBNx    \undefined \def \showISBNx     #1{\unskip}     \fi
\ifx \showISBNxiii \undefined \def \showISBNxiii  #1{\unskip}     \fi
\ifx \showISSN     \undefined \def \showISSN      #1{\unskip}     \fi
\ifx \showLCCN     \undefined \def \showLCCN      #1{\unskip}     \fi
\ifx \shownote     \undefined \def \shownote      #1{#1}          \fi
\ifx \showarticletitle \undefined \def \showarticletitle #1{#1}   \fi
\ifx \showURL      \undefined \def \showURL       {\relax}        \fi
\providecommand\bibfield[2]{#2}
\providecommand\bibinfo[2]{#2}
\providecommand\natexlab[1]{#1}
\providecommand\showeprint[2][]{arXiv:#2}

\bibitem[\protect\citeauthoryear{Bauer, Cuoq, and Regehr}{Bauer
  et~al\mbox{.}}{2015}]%
        {regehr:backdoors15}
\bibfield{author}{\bibinfo{person}{Scott Bauer}, \bibinfo{person}{Pascal Cuoq},
  {and} \bibinfo{person}{John Regehr}.} \bibinfo{year}{2015}\natexlab{}.
\newblock \showarticletitle{Deniable Backdoors using Compiler Bugs}.
\newblock \bibinfo{journal}{\emph{PoC GTFO}} (\bibinfo{year}{2015}),
  \bibinfo{pages}{7--9}.
\newblock


\bibitem[\protect\citeauthoryear{Boujarwah and Saleh}{Boujarwah and
  Saleh}{1997}]%
        {Boujarwah97}
\bibfield{author}{\bibinfo{person}{Abdulazeez Boujarwah} {and}
  \bibinfo{person}{Kassem Saleh}.} \bibinfo{year}{1997}\natexlab{}.
\newblock \showarticletitle{Compiler test case generation methods: a survey and
  assessment}.
\newblock \bibinfo{journal}{\emph{Information and Software Technology (IST)}}
  \bibinfo{volume}{39} (\bibinfo{year}{1997}), \bibinfo{pages}{617 -- 625}.
\newblock
Issue 9.


\bibitem[\protect\citeauthoryear{Burgess and Saidi}{Burgess and Saidi}{1996}]%
        {compiler-bugs:fortran}
\bibfield{author}{\bibinfo{person}{Colin Burgess} {and} \bibinfo{person}{M.
  Saidi}.} \bibinfo{year}{1996}\natexlab{}.
\newblock \showarticletitle{The automatic generation of test cases for
  optimizing {Fortran} compilers}.
\newblock \bibinfo{journal}{\emph{Information and Software Technology (IST)}}
  \bibinfo{volume}{38} (\bibinfo{year}{1996}), \bibinfo{pages}{111 -- 119}.
\newblock
Issue 2.


\bibitem[\protect\citeauthoryear{Cadar, Pina, and Regehr}{Cadar
  et~al\mbox{.}}{2015}]%
        {compiler-backdoor:blog15}
\bibfield{author}{\bibinfo{person}{Cristian Cadar},
  \bibinfo{person}{Lu{\'{\i}}s Pina}, {and} \bibinfo{person}{John Regehr}.}
  \bibinfo{year}{2015}\natexlab{}.
\newblock \bibinfo{title}{Multi-Version Execution Defeats a Compiler-Bug-Based
  Backdoor}.
\newblock
  \bibinfo{howpublished}{\url{http://ccadar.blogspot.co.uk/2015/11/multi-version-execution-defeats.html}}.
\newblock


\bibitem[\protect\citeauthoryear{Cadar and Sen}{Cadar and Sen}{2013}]%
        {symex:cacm}
\bibfield{author}{\bibinfo{person}{Cristian Cadar} {and}
  \bibinfo{person}{Koushik Sen}.} \bibinfo{year}{2013}\natexlab{}.
\newblock \showarticletitle{{Symbolic} {E}xecution for {S}oftware {T}esting:
  {T}hree {D}ecades {L}ater}.
\newblock \bibinfo{journal}{\emph{Communications of the Association for
  Computing Machinery (CACM)}} \bibinfo{volume}{56}, \bibinfo{number}{2}
  (\bibinfo{year}{2013}), \bibinfo{pages}{82--90}.
\newblock


\bibitem[\protect\citeauthoryear{Chen, Cheung, and Yiu}{Chen
  et~al\mbox{.}}{1998}]%
        {metamorphic98}
\bibfield{author}{\bibinfo{person}{T.Y. Chen}, \bibinfo{person}{S.C. Cheung},
  {and} \bibinfo{person}{S.M. Yiu}.} \bibinfo{year}{1998}\natexlab{}.
\newblock \bibinfo{booktitle}{\emph{Metamorphic testing: a new approach for
  generating next test cases}}.
\newblock \bibinfo{type}{{T}echnical {R}eport} HKUST-CS98-01.
  \bibinfo{institution}{Hong Kong University of Science and Technology}.
\newblock


\bibitem[\protect\citeauthoryear{Chen, Groce, Zhang, Wong, Fern, Eide, and
  Regehr}{Chen et~al\mbox{.}}{2013}]%
        {chen2013}
\bibfield{author}{\bibinfo{person}{Yang Chen}, \bibinfo{person}{Alex Groce},
  \bibinfo{person}{Chaoqiang Zhang}, \bibinfo{person}{Weng-Keen Wong},
  \bibinfo{person}{Xiaoli Fern}, \bibinfo{person}{Eric Eide}, {and}
  \bibinfo{person}{John Regehr}.} \bibinfo{year}{2013}\natexlab{}.
\newblock \showarticletitle{Taming Compiler Fuzzers}. In
  \bibinfo{booktitle}{\emph{Proc. of the Conference on Programing Language
  Design and Implementation (PLDI'13)}}.
\newblock


\bibitem[\protect\citeauthoryear{Chen, Su, Sun, Su, and Zhao}{Chen
  et~al\mbox{.}}{2016}]%
        {Chen2016}
\bibfield{author}{\bibinfo{person}{Yuting Chen}, \bibinfo{person}{Ting Su},
  \bibinfo{person}{Chengnian Sun}, \bibinfo{person}{Zhendong Su}, {and}
  \bibinfo{person}{Jianjun Zhao}.} \bibinfo{year}{2016}\natexlab{}.
\newblock \showarticletitle{{Coverage-directed differential testing of {JVM}
  implementations}}. In \bibinfo{booktitle}{\emph{Proc. of the Conference on
  Programing Language Design and Implementation (PLDI'16)}}.
\newblock


\bibitem[\protect\citeauthoryear{Cuoq, Monate, Pacalet, Prevosto, Regehr,
  Yakobowski, and Yang}{Cuoq et~al\mbox{.}}{2012}]%
        {csmithFramaC}
\bibfield{author}{\bibinfo{person}{Pascal Cuoq}, \bibinfo{person}{Benjamin
  Monate}, \bibinfo{person}{Anne Pacalet}, \bibinfo{person}{Virgile Prevosto},
  \bibinfo{person}{John Regehr}, \bibinfo{person}{Boris Yakobowski}, {and}
  \bibinfo{person}{Xuejun Yang}.} \bibinfo{year}{2012}\natexlab{}.
\newblock \showarticletitle{Testing Static Analyzers with Randomly Generated
  Programs}. In \bibinfo{booktitle}{\emph{Proc. of the 4th International
  Conference on NASA Formal Methods}}.
\newblock


\bibitem[\protect\citeauthoryear{Daniel, Dig, Garcia, and Marinov}{Daniel
  et~al\mbox{.}}{2007}]%
        {testing-refactorings}
\bibfield{author}{\bibinfo{person}{Brett Daniel}, \bibinfo{person}{Danny Dig},
  \bibinfo{person}{Kely Garcia}, {and} \bibinfo{person}{Darko Marinov}.}
  \bibinfo{year}{2007}\natexlab{}.
\newblock \showarticletitle{Automated Testing of Refactoring Engines}. In
  \bibinfo{booktitle}{\emph{Proc. of the joint meeting of the European Software
  Engineering Conference and the {ACM} Symposium on the Foundations of Software
  Engineering (ESEC/FSE'07)}}.
\newblock


\bibitem[\protect\citeauthoryear{Donaldson, Evrard, Lascu, and
  Thomson}{Donaldson et~al\mbox{.}}{2017}]%
        {compiler-bugs:PACMPL17}
\bibfield{author}{\bibinfo{person}{Alastair~F. Donaldson},
  \bibinfo{person}{Hugues Evrard}, \bibinfo{person}{Andrei Lascu}, {and}
  \bibinfo{person}{Paul Thomson}.} \bibinfo{year}{2017}\natexlab{}.
\newblock \showarticletitle{Automated Testing of Graphics Shader Compilers}.
\newblock \bibinfo{journal}{\emph{Proceedings of the ACM Programming Languages
  (PACMPL)}} \bibinfo{volume}{1}, \bibinfo{number}{{OOPSLA}}
  (\bibinfo{year}{2017}), \bibinfo{pages}{93:1--93:29}.
\newblock


\bibitem[\protect\citeauthoryear{Donaldson and Lascu}{Donaldson and
  Lascu}{2016}]%
        {OpenGLtesting:met16}
\bibfield{author}{\bibinfo{person}{Alastair~F. Donaldson} {and}
  \bibinfo{person}{Andrei Lascu}.} \bibinfo{year}{2016}\natexlab{}.
\newblock \showarticletitle{Metamorphic testing for (graphics) compilers}. In
  \bibinfo{booktitle}{\emph{Proc. of the International Workshop on Metamorphic
  Testing (MET'16)}}.
\newblock


\bibitem[\protect\citeauthoryear{Hanford}{Hanford}{1970}]%
        {Hanford70}
\bibfield{author}{\bibinfo{person}{K.V. Hanford}.}
  \bibinfo{year}{1970}\natexlab{}.
\newblock \showarticletitle{{Automatic generation of test cases}}.
\newblock \bibinfo{journal}{\emph{IBM Systems Journal}}  \bibinfo{volume}{9}
  (\bibinfo{year}{1970}), \bibinfo{pages}{242--257}.
\newblock
Issue 4.


\bibitem[\protect\citeauthoryear{Holler, Herzig, and Zeller}{Holler
  et~al\mbox{.}}{2012}]%
        {fuzzing-fragments}
\bibfield{author}{\bibinfo{person}{Christian Holler}, \bibinfo{person}{Kim
  Herzig}, {and} \bibinfo{person}{Andreas Zeller}.}
  \bibinfo{year}{2012}\natexlab{}.
\newblock \showarticletitle{Fuzzing with Code Fragments}. In
  \bibinfo{booktitle}{\emph{Proc. of the 21st {USENIX} Security Symposium
  ({USENIX} Security'12)}}.
\newblock


\bibitem[\protect\citeauthoryear{Hosek and Cadar}{Hosek and Cadar}{2015}]%
        {varan}
\bibfield{author}{\bibinfo{person}{Petr Hosek} {and} \bibinfo{person}{Cristian
  Cadar}.} \bibinfo{year}{2015}\natexlab{}.
\newblock \showarticletitle{{Varan the Unbelievable}: An efficient {N}-version
  execution framework}. In \bibinfo{booktitle}{\emph{Proc. of the 20th
  International Conference on Architectural Support for Programming Languages
  and Operating Systems (ASPLOS'15)}}.
\newblock


\bibitem[\protect\citeauthoryear{Jones}{Jones}{2015}]%
        {anti-comp-fuzzing15}
\bibfield{author}{\bibinfo{person}{Derek Jones}.}
  \bibinfo{year}{2015}\natexlab{}.
\newblock \bibinfo{title}{So you found a bug in my compiler: Whoopee do}.
\newblock
  \bibinfo{howpublished}{\url{http://shape-of-code.coding-guidelines.com/2015/12/07/so-you-found-a-bug-in-my-compiler-whoopee-do/}}.
\newblock


\bibitem[\protect\citeauthoryear{Kapus and Cadar}{Kapus and Cadar}{2017}]%
        {symex-test}
\bibfield{author}{\bibinfo{person}{Timotej Kapus} {and}
  \bibinfo{person}{Cristian Cadar}.} \bibinfo{year}{2017}\natexlab{}.
\newblock \showarticletitle{Automatic Testing of Symbolic Execution Engines via
  Program Generation and Differential Testing}. In
  \bibinfo{booktitle}{\emph{Proc. of the 32nd IEEE International Conference on
  Automated Software Engineering (ASE'17)}}.
\newblock


\bibitem[\protect\citeauthoryear{Kossatchev and Posypkin}{Kossatchev and
  Posypkin}{2005}]%
        {Kossatchev2005}
\bibfield{author}{\bibinfo{person}{Alexander Kossatchev} {and}
  \bibinfo{person}{Mikhail Posypkin}.} \bibinfo{year}{2005}\natexlab{}.
\newblock \showarticletitle{Survey of Compiler Testing Methods}.
\newblock \bibinfo{journal}{\emph{Programming and Computing Software}}
  \bibinfo{volume}{31} (\bibinfo{date}{Jan.} \bibinfo{year}{2005}),
  \bibinfo{pages}{10--19}.
\newblock
Issue 1.


\bibitem[\protect\citeauthoryear{Lattner and Adve}{Lattner and Adve}{2004}]%
        {llvm}
\bibfield{author}{\bibinfo{person}{Chris Lattner} {and} \bibinfo{person}{Vikram
  Adve}.} \bibinfo{year}{2004}\natexlab{}.
\newblock \showarticletitle{{LLVM}: A Compilation Framework for Lifelong
  Program Analysis \& Transformation}. In \bibinfo{booktitle}{\emph{Proc. of
  the 2nd International Symposium on Code Generation and Optimization
  (CGO'04)}}.
\newblock


\bibitem[\protect\citeauthoryear{Le, Afshari, and Su}{Le et~al\mbox{.}}{2014}]%
        {emi}
\bibfield{author}{\bibinfo{person}{Vu Le}, \bibinfo{person}{Mehrdad Afshari},
  {and} \bibinfo{person}{Zhendong Su}.} \bibinfo{year}{2014}\natexlab{}.
\newblock \showarticletitle{Compiler Validation via Equivalence Modulo Inputs}.
  In \bibinfo{booktitle}{\emph{Proc. of the Conference on Programing Language
  Design and Implementation (PLDI'14)}}.
\newblock


\bibitem[\protect\citeauthoryear{Le, Sun, and Su}{Le et~al\mbox{.}}{2015a}]%
        {athena}
\bibfield{author}{\bibinfo{person}{Vu Le}, \bibinfo{person}{Chengnian Sun},
  {and} \bibinfo{person}{Zhendong Su}.} \bibinfo{year}{2015}\natexlab{a}.
\newblock \showarticletitle{Finding Deep Compiler Bugs via Guided Stochastic
  Program Mutation}. In \bibinfo{booktitle}{\emph{Proc. of the 30th Annual
  Conference on Object-Oriented Programming Systems, Languages and Applications
  (OOPSLA'15)}}.
\newblock


\bibitem[\protect\citeauthoryear{Le, Sun, and Su}{Le et~al\mbox{.}}{2015b}]%
        {le2015issta}
\bibfield{author}{\bibinfo{person}{Vu Le}, \bibinfo{person}{Chengnian Sun},
  {and} \bibinfo{person}{Zhendong Su}.} \bibinfo{year}{2015}\natexlab{b}.
\newblock \showarticletitle{Randomized Stress-testing of Link-time Optimizers}.
  In \bibinfo{booktitle}{\emph{Proc. of the International Symposium on Software
  Testing and Analysis (ISSTA'15)}}.
\newblock


\bibitem[\protect\citeauthoryear{Leroy}{Leroy}{2009}]%
        {CompCert}
\bibfield{author}{\bibinfo{person}{Xavier Leroy}.}
  \bibinfo{year}{2009}\natexlab{}.
\newblock \showarticletitle{Formal verification of a realistic compiler}.
\newblock \bibinfo{journal}{\emph{Communications of the Association for
  Computing Machinery (CACM)}} \bibinfo{volume}{52}, \bibinfo{number}{7}
  (\bibinfo{year}{2009}), \bibinfo{pages}{107--115}.
\newblock


\bibitem[\protect\citeauthoryear{Lidbury, Lascu, Chong, and Donaldson}{Lidbury
  et~al\mbox{.}}{2015}]%
        {clsmith}
\bibfield{author}{\bibinfo{person}{Christopher Lidbury},
  \bibinfo{person}{Andrei Lascu}, \bibinfo{person}{Nathan Chong}, {and}
  \bibinfo{person}{Alastair~F. Donaldson}.} \bibinfo{year}{2015}\natexlab{}.
\newblock \showarticletitle{Many-core compiler fuzzing}. In
  \bibinfo{booktitle}{\emph{Proc. of the Conference on Programing Language
  Design and Implementation (PLDI'15)}}.
\newblock


\bibitem[\protect\citeauthoryear{Lopes, Menendez, Nagarakatte, and
  Regehr}{Lopes et~al\mbox{.}}{2015}]%
        {alive}
\bibfield{author}{\bibinfo{person}{Nuno Lopes}, \bibinfo{person}{David
  Menendez}, \bibinfo{person}{Santosh Nagarakatte}, {and} \bibinfo{person}{John
  Regehr}.} \bibinfo{year}{2015}\natexlab{}.
\newblock \showarticletitle{Provably Correct Peephole Optimizations with
  Alive}. In \bibinfo{booktitle}{\emph{Proc. of the Conference on Programing
  Language Design and Implementation (PLDI'15)}}.
\newblock


\bibitem[\protect\citeauthoryear{Luo, Hariri, Eloussi, and Marinov}{Luo
  et~al\mbox{.}}{2014}]%
        {flaky:fse14}
\bibfield{author}{\bibinfo{person}{Qingzhou Luo}, \bibinfo{person}{Farah
  Hariri}, \bibinfo{person}{Lamyaa Eloussi}, {and} \bibinfo{person}{Darko
  Marinov}.} \bibinfo{year}{2014}\natexlab{}.
\newblock \showarticletitle{An Empirical Analysis of Flaky Tests}. In
  \bibinfo{booktitle}{\emph{Proc. of the {ACM} {SIGSOFT} Symposium on the
  Foundations of Software Engineering (FSE'14)}}.
\newblock


\bibitem[\protect\citeauthoryear{Marinescu, Hosek, and Cadar}{Marinescu
  et~al\mbox{.}}{2014}]%
        {covrig}
\bibfield{author}{\bibinfo{person}{Paul~Dan Marinescu}, \bibinfo{person}{Petr
  Hosek}, {and} \bibinfo{person}{Cristian Cadar}.}
  \bibinfo{year}{2014}\natexlab{}.
\newblock \showarticletitle{Covrig: A Framework for the Analysis of Code, Test,
  and Coverage Evolution in Real Software}. In \bibinfo{booktitle}{\emph{Proc.
  of the International Symposium on Software Testing and Analysis (ISSTA'14)}}.
\newblock


\bibitem[\protect\citeauthoryear{McKeeman}{McKeeman}{1998}]%
        {mckeeman:diff-test}
\bibfield{author}{\bibinfo{person}{W.~M. McKeeman}.}
  \bibinfo{year}{1998}\natexlab{}.
\newblock \showarticletitle{Differential testing for software}.
\newblock \bibinfo{journal}{\emph{Digital Technical Journal}}
  \bibinfo{volume}{10} (\bibinfo{year}{1998}), \bibinfo{pages}{100--107}.
\newblock
Issue 1.


\bibitem[\protect\citeauthoryear{Nagai, Hashimoto, and Ishiura}{Nagai
  et~al\mbox{.}}{2014}]%
        {orange3}
\bibfield{author}{\bibinfo{person}{Eriko Nagai}, \bibinfo{person}{Atsushi
  Hashimoto}, {and} \bibinfo{person}{Nagisa Ishiura}.}
  \bibinfo{year}{2014}\natexlab{}.
\newblock \showarticletitle{Reinforcing random testing of arithmetic
  optimization of C compilers by scaling up size and number of expressions}.
\newblock \bibinfo{journal}{\emph{IPSJ Transactions on System LSI Design
  Methodology}}  \bibinfo{volume}{7} (\bibinfo{year}{2014}),
  \bibinfo{pages}{91--100}.
\newblock


\bibitem[\protect\citeauthoryear{Nakamura and Ishiura}{Nakamura and
  Ishiura}{2016}]%
        {orange4}
\bibfield{author}{\bibinfo{person}{Kazuhiro Nakamura} {and}
  \bibinfo{person}{Nagisa Ishiura}.} \bibinfo{year}{2016}\natexlab{}.
\newblock \showarticletitle{Random testing of C compilers based on test program
  generation by equivalence transformation}. In \bibinfo{booktitle}{\emph{2016
  IEEE Asia Pacific Conference on Circuits and Systems (APCCAS)}}.
  \bibinfo{pages}{676--679}.
\newblock


\bibitem[\protect\citeauthoryear{Purdom}{Purdom}{1972}]%
        {Purdom72}
\bibfield{author}{\bibinfo{person}{Paul Purdom}.}
  \bibinfo{year}{1972}\natexlab{}.
\newblock \showarticletitle{A sentence generator for testing parsers}.
\newblock \bibinfo{journal}{\emph{BIT Numerical Mathematics}}
  \bibinfo{volume}{12} (\bibinfo{year}{1972}), \bibinfo{pages}{366--375}.
\newblock
Issue 3.


\bibitem[\protect\citeauthoryear{Regehr, Chen, Cuoq, Eide, Ellison, and
  Yang}{Regehr et~al\mbox{.}}{2012}]%
        {creduce}
\bibfield{author}{\bibinfo{person}{John Regehr}, \bibinfo{person}{Yang Chen},
  \bibinfo{person}{Pascal Cuoq}, \bibinfo{person}{Eric Eide},
  \bibinfo{person}{Chucky Ellison}, {and} \bibinfo{person}{Xuejun Yang}.}
  \bibinfo{year}{2012}\natexlab{}.
\newblock \showarticletitle{Test-case reduction for {C} compiler bugs}. In
  \bibinfo{booktitle}{\emph{Proc. of the Conference on Programing Language
  Design and Implementation (PLDI'12)}}.
\newblock


\bibitem[\protect\citeauthoryear{Sauder}{Sauder}{1962}]%
        {cobol-testing62}
\bibfield{author}{\bibinfo{person}{Richard~L. Sauder}.}
  \bibinfo{year}{1962}\natexlab{}.
\newblock \showarticletitle{A General Test Data Generator for {COBOL}}. In
  \bibinfo{booktitle}{\emph{Proc. of the 1962 Spring Joint Computer Conference
  (AIEE-IRE'62 Spring)}}.
\newblock


\bibitem[\protect\citeauthoryear{Segura, Fraser, Sanchez, and
  Ruiz-Cort{\'e}s}{Segura et~al\mbox{.}}{2016}]%
        {segura2016survey}
\bibfield{author}{\bibinfo{person}{Sergio Segura}, \bibinfo{person}{Gordon
  Fraser}, \bibinfo{person}{Ana Sanchez}, {and} \bibinfo{person}{Antonio
  Ruiz-Cort{\'e}s}.} \bibinfo{year}{2016}\natexlab{}.
\newblock \showarticletitle{A Survey on Metamorphic Testing}.
\newblock  (\bibinfo{year}{2016}).
\newblock


\bibitem[\protect\citeauthoryear{Sun, Le, and Su}{Sun et~al\mbox{.}}{2016a}]%
        {hermes}
\bibfield{author}{\bibinfo{person}{Chengnian Sun}, \bibinfo{person}{Vu Le},
  {and} \bibinfo{person}{Zhendong Su}.} \bibinfo{year}{2016}\natexlab{a}.
\newblock \showarticletitle{Finding compiler bugs via live code mutation}. In
  \bibinfo{booktitle}{\emph{Proc. of the 31st Annual Conference on
  Object-Oriented Programming Systems, Languages and Applications
  (OOPSLA'16)}}.
\newblock


\bibitem[\protect\citeauthoryear{Sun, Le, Zhang, and Su}{Sun
  et~al\mbox{.}}{2016b}]%
        {compiler-bugs:issta16}
\bibfield{author}{\bibinfo{person}{Chengnian Sun}, \bibinfo{person}{Vu Le},
  \bibinfo{person}{Qirun Zhang}, {and} \bibinfo{person}{Zhendong Su}.}
  \bibinfo{year}{2016}\natexlab{b}.
\newblock \showarticletitle{Toward Understanding Compiler Bugs in GCC and
  LLVM}. In \bibinfo{booktitle}{\emph{Proc. of the International Symposium on
  Software Testing and Analysis (ISSTA'16)}}.
\newblock


\bibitem[\protect\citeauthoryear{Tao, Wu, Zhao, and Shen}{Tao
  et~al\mbox{.}}{2010}]%
        {compiler-bugs:metamorphic-testing}
\bibfield{author}{\bibinfo{person}{Qiuming Tao}, \bibinfo{person}{Wei Wu},
  \bibinfo{person}{Chen Zhao}, {and} \bibinfo{person}{Wuwei Shen}.}
  \bibinfo{year}{2010}\natexlab{}.
\newblock \showarticletitle{An Automatic Testing Approach for Compiler Based on
  Metamorphic Testing Technique}. In \bibinfo{booktitle}{\emph{Proc. of the
  17th Asia-Pacific Software Engineering Conference (ASPEC'10)}}.
\newblock


\bibitem[\protect\citeauthoryear{Wichmann}{Wichmann}{1998}]%
        {Wichmann98}
\bibfield{author}{\bibinfo{person}{B.A. Wichmann}.}
  \bibinfo{year}{1998}\natexlab{}.
\newblock \bibinfo{title}{Some Remarks about Random Testing}.
\newblock
\newblock
\newblock
\shownote{\url{http://www.npl.co.uk/upload/pdf/random_testing.pdf}.}


\bibitem[\protect\citeauthoryear{Yang, Chen, Eide, and Regehr}{Yang
  et~al\mbox{.}}{2011}]%
        {csmith}
\bibfield{author}{\bibinfo{person}{Xuejun Yang}, \bibinfo{person}{Yang Chen},
  \bibinfo{person}{Eric Eide}, {and} \bibinfo{person}{John Regehr}.}
  \bibinfo{year}{2011}\natexlab{}.
\newblock \showarticletitle{Finding and Understanding Bugs in {C} Compilers}.
  In \bibinfo{booktitle}{\emph{Proc. of the Conference on Programing Language
  Design and Implementation (PLDI'11)}}.
\newblock


\bibitem[\protect\citeauthoryear{Yarpgen}{Yarpgen}{2018}]%
        {yarpgen}
\bibfield{author}{\bibinfo{person}{Yarpgen}.} \bibinfo{year}{2018}\natexlab{}.
\newblock \bibinfo{title}{{\url{https://github.com/intel/yarpgen}}}.
\newblock
\newblock


\bibitem[\protect\citeauthoryear{Zhang, Sun, and Su}{Zhang
  et~al\mbox{.}}{2017}]%
        {skeletal-program-enum}
\bibfield{author}{\bibinfo{person}{Qirun Zhang}, \bibinfo{person}{Chengnian
  Sun}, {and} \bibinfo{person}{Zhendong Su}.} \bibinfo{year}{2017}\natexlab{}.
\newblock \showarticletitle{Skeletal program enumeration for rigorous compiler
  testing}. In \bibinfo{booktitle}{\emph{Proc. of the Conference on Programing
  Language Design and Implementation (PLDI'17)}}.
\newblock


\end{thebibliography}
